\documentclass[sigconf]{acmartarxiv}

\def\BibTeX{{\rm B\kern-.05em{\sc i\kern-.025em b}\kern-.08emT\kern-.1667em\lower.7ex\hbox{E}\kern-.125emX}}
    
\newif\ifPATENT
\newif\ifICPP
\newif\ifARXIV

\ARXIVtrue
\ifARXIV
	\setcopyright{none}
\fi

\ICPPtrue
\PATENTfalse

\usepackage{graphicx,pifont,color} 

\usepackage{multirow}
\usepackage{paralist}
\usepackage[final]{pdfpages}

\usepackage{lhelp}
\usepackage{xcolor,colortbl}

\definecolor{wp}{rgb}{0.129,0.478, 0.678}
\definecolor{wpbright}{rgb}{0.26,0.55, 0.79}
\definecolor{yellowBright}{HTML}{FFD247}
\definecolor{orangeBright}{HTML}{FF7E47}
\definecolor{gray1}{HTML}{CEDBE1}
\definecolor{gray2}{HTML}{DDE6EA}
\definecolor{gray3}{HTML}{E3E6EB}
\definecolor{darkblue1}{HTML}{2375B9}
\definecolor{darkblue2}{HTML}{3D577A}
\definecolor{lightblue1}{HTML}{65A4D9}
\definecolor{lightblue2}{HTML}{94C3E9}
\definecolor{red1}{HTML}{E4001B}
\definecolor{red2}{HTML}{8F2831}
\definecolor{lime1}{HTML}{AEBD63}
\definecolor{navy}{HTML}{000063}
\definecolor{purple1}{HTML}{8F2831}
\definecolor{contrast}{HTML}{CE0000}
\definecolor{positive}{HTML}{0D900F}
\definecolor{negative}{HTML}{FF0000}

\newcolumntype{a}{>{\columncolor{gray1}}c}

\makeatletter
\DeclareRobustCommand*\cal{\@fontswitch\relax\mathcal}
\makeatother

\newif\ifready
\readyfalse

\newif\ifstable
\stabletrue

\newif\ifnewversion
\newversiontrue

\def\cuda{\mbox{\sc CUDA}}

\def\mwpcwp{\mbox{MWP-CWP}}

\newcommand{\blockcomment}[1]{}

\newtheorem{definition}{Definition}

\newtheorem{example}{Example}

\newtheorem{observation}{Observation}
\newtheorem{remark}{Remark}
\newtheorem{consequence}{Consequence}

\newcommand{\hidetext}[1]{\mbox{ \ }}
\newcommand{\todo}[2]{{{\textcolor{red}{ #1}}}\footnote{ {\textcolor{blue}{ #2}} }}
\newcommand{\fixed}[2]{#1}

\def\Q{\mbox{${\mathbb Q}$}}

\newcommand{\nvprof}{\texttt{nvprof}}

\title{A Technique for Finding Optimal Program Launch Parameters Targeting
        Manycore Accelerators}
\author{Alexander Brandt}
\affiliation{\institution{\small University of Western Ontario, Canada}}
\email{abrandt5@uwo.ca}

\author{Davood Mohajerani}
\affiliation{\institution{\small University of Western Ontario, Canada}}
\email{dmohajer@uwo.ca}

\author{Marc~Moreno~Maza}
\affiliation{\institution{\small University of Western Ontario, Canada}}
\email{moreno@csd.uwo.ca}

\author{Jeeva Paudel}
\affiliation{\institution{\small IBM Canada Software Laboratory, Markham, Canada}}
\email{pjeeva01@ca.ibm.com}

\author{Lin-Xiao Wang}
\affiliation{\institution{\small University of Western Ontario, Canada}}
\email{lwang739@uwo.ca}

\renewcommand{\shortauthors}{Brandt, Mohajerani, Moreno~Maza, Paudel, Wang}

\begin{document}

\begin{abstract}
\ifPATENT
In this disclosure we have presented a new technique to dynamically
determine the program parameters which optimize the performance of a
multithreaded program.  To be precise, we have described \textit{a
novel technique to statically build a program that can dynamically
determine the optimal values of program parameters to yield the best
program performance for given data and hardware parameters}.
\fi
\ifICPP
In this paper, we present a new technique to dynamically determine the
values of program parameters in order to optimize the performance of a
multithreaded program ${\cal P}$. To be precise, we describe a novel
technique to statically build another program, say, ${\cal R}$, 
 that can dynamically determine
the optimal values of program parameters to yield the best program
performance for ${\cal P}$ given values for its data and hardware parameters.
\fi 
\ifPATENT

A program's performance is typically quantified by some high-level
metric such as execution time, memory consumption or hardware
occupancy. We statically build a \textit{rational program} (see
Section~\ref{sec:TheIdea}) to dynamically estimate such a high-level
metric from the values of data and hardware parameters.  As shown in
Section~\ref{sect:steps}, this program is built from the following
three steps:
\begin{inparaenum}[(i)]
	\item emulation of the multithreaded program to collect statistics of low-level metrics,
	\item determine rational functions to estimate these low-level metrics, and
	\item generate a rational program containing these rational functions to estimate the high-level metric.
\end{inparaenum}

Once the rational program has been constructed, we use it to determine
the optimal program parameters of the multithreaded program.  By
giving the data, hardware, and program parameters particular values in
the rational program, it produces an estimate for the high-level
metric. At runtime, where the data and hardware parameters are known,
this allows us to perform an exhaustive search over the possible program
parameters to determine which configurations optimize the high-level
metric for the given data and hardware parameters.

An important fact of this process is that the rational program
is built {\bf completely off-line}, that is, at compile-time. 
Once the rational program is constructed it can be used
dynamically at runtime to optimize the program for
the present data and hardware parameters. This allows users
to optimally and easily execute the same parallel code
on different hardware platforms and with different input data
sizes.

\fi
While this technique can be applied to parallel programs in general, 
we are particularly interested in programs targeting manycore accelerators.
Our technique has successfully been employed for GPU kernels using the
\mwpcwp\ performance model for \cuda. 
\ifPATENT
Sections~\ref{sec:embodiment} and
\ref{sec:example} have shown how one can apply our general technique
to programs specifically targeting GPU architectures,
while Section~\ref{sec:performance} has given empirical data
showing the success of this realization.
\fi

\end{abstract}

%
%

\maketitle

\section{Introduction}

Three types of parameters influence the performance of parallel programs on multiprocessors:
  \begin{inparaenum}[(i)]
  \item {\em data parameters}, such as input data and its size,  
  \item {\em hardware parameters}, such as cache capacity and number of 
         available registers, and
  \item {\em program parameters}, such as granularity of tasks and 
        the quantities that characterize how tasks are
        mapped to processors.
  \end{inparaenum}

  Data and hardware parameters are independent from program parameters,
  and are determined by users' needs and available hardware resources.
  Program parameters, however, are intimately related to data and
  hardware parameters. Meanwhile, the choice of program parameters can largely 
  effect the performance of the program. Therefore, determining optimal values of
  program parameters that yield the best program performance for a given
  confluence of hardware and data parameter values is critical. Further,
  determining such values automatically is important to enable users
  to execute the same parallel program efficiently on different
  hardware platforms.

\ifPATENT  
  This disclosure describes {\em a novel technique to statically build
    a program that can dynamically determine the optimal values of
    program parameters to yield the best program performance for given
    data and hardware parameters}.

\fi
\ifICPP
  This paper describes a novel technique to statically build
    a program ${\cal R}$ that can dynamically determine the optimal values of
    program parameters to yield the best program performance for given
    values of the data and hardware parameters of a given multithreaded
    program ${\cal P}$.
\fi
  The key principle underpinning the proposed technique can be
  summarized as follows.  
\ifICPP

\fi 
In most program execution models, high-level performance metrics, such
as execution time, memory consumption, and hardware occupancy, are
piece-wise rational functions (PRFs) of lower-level metrics, which
include the number of cache misses and the number of cycles per
instruction.  These lower-level metrics are themselves PRFs of
program, hardware, and data parameters. As such, for a fixed machine,
a high-level performance metric can be estimated by a piece-wise
rational function of the program and data parameters. Henceforth, we
regard a computer program that computes such a piece-wise rational
function as a special sort of {\em rational program},
\ifPATENT
a technical notion defined formally below.
\fi
\ifICPP
a technical notion defined in Section~\ref{sect:ratprog}.
\fi

\ifICPP
\subsection{Problem Statement}
\fi
Let ${\cal P}$ be a multithreaded program to be executed on a targeted
multiprocessor. 
By fixing the target architecture, the hardware
parameters, say, $H_1, \ldots, H_h$ then become fixed and 
we can assume that the performance of
${\cal P}$ depends only on data parameters $D_1, \ldots, D_d$ and program
parameters $P_1, \ldots, P_p$. Also, the optimal choice of $P_1, \ldots, P_p$ depends on a specific choice of $D_1, \ldots, D_d$.
For example, in programs targeting GPUs 
(e.g. programs written in {\cuda}),   
the parameters $D_1, \ldots, D_d$ are typically
dimension sizes of data structures, like arrays, 
while $P_1, \ldots, P_p$ typically specify the formats
of thread blocks (see Section~\ref{sec:embodiment} for further details of 
this technique applied to \cuda).

In most cases, data parameters are only given at runtime, 
which makes it difficult to determine the optimal program 
parameters before that. However, a bad choice of program 
parameters can be very inefficient, especially for programs 
that consume a large amount of resources 
(such as running time and memory consumption). 
Hence, it is crucial to be able to determine the optimal program 
parameters at runtime without much overhead added to the program 
execution.

\ifPATENT
Hence, the parameters do not include the actual input data.
This is not, however, a limitation or restriction of our technique.
It is necessary only to \textit{describe} the data, say by its size or
structure, in order to characterize its impact on performance, rather
than needing the data itself.
\fixed{In fact, the program ${\cal P}$ should be understood as}{I find this section does not add to the discussion at all.
JP: To me, the two bulleted points in fact substantiate the previous statement about the importance of the characteristics of the input data
rather than the availability of the actual data itself.}
\begin{itemizeshort}
\item the implementation of an algorithm working with dense
      data, e.g. in dense linear algebra or in signal processing, or
\item the implementation of an algorithm working with structured sparse
      data, e.g. graph colouring for a particular family of graphs, 
      say the Generalized Petersen Graphs
      $GPG(n,p)$.
\end{itemizeshort}
\fi
Let ${\cal E}$ be a performance metric for ${\cal P}$ that we want to
optimize.  More precisely, given the values of the data parameters
$D_1, \ldots, D_d$, the goal is to find values of the program parameters
$P_1, \ldots, P_p$ such that the execution of ${\cal P}$ optimizes
${\cal E}$.  Here, optimizing could mean maximizing, as in the case of a
performance metric such as hardware occupancy, or minimizing, as in the case
of a performance metric like execution time. 
To address our goal, we compute a mathematical expression, parameterized by data and program parameters,
in the format of a {\em rational program} ${\cal R}$ (see Section~\ref{sect:ratprog}) at compile-time. 
At runtime, given the specific values of $D_1, \ldots, D_d$, we can efficiently evaluate ${\cal E}$ 
using ${\cal R}$.
After that, we can feasibly pick the $P_1, \ldots, P_p$ that optimize ${\cal E}$, and feed that to ${\cal P}$ for execution.

\ifICPP
\subsection{Contributions}
Towards the aforementioned goal, our specific contributions are:
\begin{enumerate}[(i)]
	\item a technique for devising a mathematical expression in the form of a {\em rational program} ${\cal R}$ (Section~\ref{sect:ratprog}) that evaluates ${\cal E}$ as a function of $D_1, \ldots, D_d$ and $P_1, \ldots, P_p$,
	\item an executable example of the rational program in the form of a C program using the MWP-CWP model~\cite{hong90}, and
	\item an empirical evaluation of this example on  kernels in the \texttt{Polybench/GPU} benchmark suite.
\end{enumerate}
\fi

\ifICPP
\subsection{Structure of the Paper}
The rest of this paper is organized as follows.
Section 2 formalizes and exemplifies the notion of rational programs.
\fi
Section~\ref{sect:keyobservation} states two observations about the
targeted rational program ${\cal R}$, from which
Section~\ref{sec:consequences} derives consequences.  Such
consequences will lead to a step-by-step description, in
Section~\ref{sect:steps}, of our proposed approach for building ${\cal
R}$.

We shall see in Section~\ref{sec:embodiment} that ${\cal R}$ can be
built when the program ${\cal P}$ is being compiled, that is,
at \textit{compile-time}.  Then, when ${\cal P}$ is executed on a
specific input data set, that is, at \textit{runtime},
the rational program ${\cal R}$ can be used
to make an optimal choice for the program parameters of ${\cal P}$.
The embodiment of the propped techniques, as reported in 
Section~\ref{sec:embodiment}, targets Graphics Processing Units (GPUs).

Section~\ref{sec:performance} presents empirical results from
  evaluation of this technique on the {\tt PolyBench} test suite.
Section~\ref{sec:relatedwork} compares and contrasts
\ifPATENT
the proposed idea with related prior art.
\fi
\ifICPP
the proposed idea with previous and related works.
\fi

  \textbf{Note:} While the idea and the technique described herein are
  applicable to parallel programs in general, this document extensively
  uses GPU architecture and the CUDA (Compute Unified Device Architecture,  see~\cite{Nickolls:2008:SPP:1365490.1365500,cuda2015})
  programming model for ease of illustration and description.

\ifPATENT
\subsection{Rational program}
\else
\section{Rational program}
\fi
\label{sect:ratprog}

Let $X_1, \ldots, X_n, Y$ be pairwise different variables~\footnote{Here
variable refers to both the algebraic meaning of a 
polynomial variable and the programming language concept.}.
Let ${\cal S}$ be a sequence  of three-address code (TAC~\cite{Aho:1986:CPT:6448})
instructions such that the set of 
the variables that occur in ${\cal S}$ 
and are never assigned a value
by an instruction of ${\cal S}$ is exactly $\{ X_1, \ldots, X_n \}$.

\begin{definition}
\label{defi:rationalprogram}
We say that the sequence ${\cal S}$ is 
{\em rational} if every
arithmetic operation used in ${\cal S}$ is either an addition, a subtraction,
a multiplication, or a  comparison ($=$, $<$), for integer numbers
either in fixed precision or in arbitrary precision.
Moreover, we say that the sequence ${\cal S}$ is 
a {\em rational program} in $X_1, \ldots, X_n$ evaluating $Y$
if the following two conditions hold:
\begin{enumerateshort}
\item ${\cal S}$ is rational, and
\item  after specializing each of $X_1, \ldots, X_n$ to an arbitrary 
      integer value in ${\cal S}$,
      the execution of the 
      specialized sequence ${\cal S}$ always terminates and the last
      executed instruction assigns an integer value to $Y$.
\end{enumerateshort}
\end{definition}

The above definition calls for a few natural remarks.

\begin{remark}
{\em
\label{remark:div}
One can easily extend Definition~\ref{defi:rationalprogram}
by allowing the use of the Euclidean division for integers, 
in both fixed and arbitrary precision.
Indeed, one can write a rational program evaluating the 
integer quotient $Q$ of a signed integer $A$ by a signed integer $B$.
Then, the remainder $R$ in the Euclidean division
of $A$ by $B$ is simply $A - Q B$.
}
\end{remark}

\begin{remark}
{\em
One can also extend Definition~\ref{defi:rationalprogram}
by allowing addition, subtraction,
multiplication and  comparison ($=$, $<$) of rational numbers
in arbitrary precision.
Indeed, each of these operations can easily be implemented
by rational programs using addition, subtraction,
multiplication and  comparison ($=$, $<$) for integer numbers
in arbitrary precision.
}
\end{remark}

\begin{remark}
\label{remark:floor}
{\em
Next, one can extend Definition~\ref{defi:rationalprogram}
by allowing the {\em integer part} operations
$F \longmapsto \lceil F \rceil$
and $F \longmapsto \lfloor F \rfloor$
where $F$ is an arbitrary rational number.
Indeed, for a rational number $A/B$ 
(where $A$ and $B$ are signed integers)
the integer parts $\lceil A/B \rceil$
and $\lfloor A/B \rfloor$ can be computed
by performing the Euclidean division
of $A$ by $B$.
Consequently, one can allow TAC instructions
of the forms $Q = \lceil A/B \rceil$
and $Q = \lfloor A/B \rfloor$, where $=$ 
denotes the assignment as in the C programming
language.
}
\end{remark}

\begin{remark}
{\em
Extending Definition~\ref{defi:rationalprogram} according to
Remarks \ref{remark:div}--\ref{remark:floor} does not change
the class of rational programs. Thus, adding Euclidean division, 
rational number arithmetic and integer part computation to the 
definition of a rational sequence yields an equivalent definition
of rational program. We adopt this definition henceforth.
}
\end{remark}

\begin{remark}
{\em
Recall that it is convenient 
to associate any sequence ${\cal S}$ of computer program 
instructions
with a {\em control flow graph} (CFG).
In the CFG of ${\cal S}$, the nodes are the {\em basic blocks} of ${\cal S}$,
that is, the sub-sequences of ${\cal S}$  such that 
\begin{inparaenum}[(i)]
  \item  each instruction except the last one is not a branch, and
   \item which are maximum in length with property $(i)$.
 \end{inparaenum}
Moreover, in the CFG of ${\cal S}$, 
there is a directed edge from a basic block $B_1$ to a basic block
$B_2$ whenever, during the execution of ${\cal S}$, one can jump from
the last instruction of $B_1$ to the first instruction of $B_2$.
Recall also that  a {\em flow chart} is another
graphic representation of a sequence of computer program  instructions.
In fact, CFGs can be seen as particular flow charts.

If, in a given flow chart ${\cal C}$, every arithmetic
operation occurring in every (process or decision) node is
either an addition, subtraction, multiplication, or comparison, for integers
in either fixed or arbitrary precision (or any other operation
as explained in Remarks~\ref{remark:div}-\ref{remark:floor}),
then ${\cal C}$ is the flow chart of a rational sequence of computer
program instructions.
Therefore, it is meaningful to depict rational programs
\ifPATENT
in the sense of Definition~\ref{defi:rationalprogram}
\fi
using flow charts. An example is given by 
\ifPATENT
Figure~\ref{fig:fig:occupancysimpleflowchart}, the meaning of which is
explained in Figure~\ref{fig:occupancysimpleflowchart}.
\fi
\ifICPP
Figure~\ref{fig:fig:occupancysimpleflowchart}.
\fi
}
\end{remark}

\input{occupancy-flowchart}

\ifPATENT

\else
\subsection{Examples}
\fi

\begin{example}
\label{ex:cuda}
{\em 
{\em Hardware occupancy}, as defined 
in the {\cuda} programming model, is a measure of how effective a program
is making use of the hardware's processors 
(Streaming Multiprocessors in case of GPUs).
Hardware occupancy is calculated from hardware parameters, namely:
\begin{itemizeshort}
\item[-] the maximum number $R_{\rm max}$ of registers per thread block,
\item[-] the maximum number $Z_{\rm max}$ of shared memory words per thread block,
\item[-] the maximum number $T_{\rm max}$ of threads per thread block,
\item[-] the maximum number $B_{\rm max}$ of thread blocks per 
      Streaming Multiprocessor (SM) and 
\item[-] the maximum number $W_{\rm max}$ of warps per SM,
\end{itemizeshort}
as well as low-level performance metrics, namely:
\begin{itemizeshort}
\item[-] the number $R$ of registers used per thread and 
\item[-] the number $Z$ of shared memory words used per thread block,
\end{itemizeshort}
and a program parameter, namely the 
number $T$ of threads per thread block.
The hardware occupancy of a {\cuda} kernel is defined as
the ratio between the number of active warps
per SM and the maximum number of warps per SM, namely
$W_{\rm active} / W_{\rm max}$, where
\ifPATENT
$W_{\rm active}$ is given by
\fi
\begin{equation}
\label{eq:occupancy}
    W_{\rm active} =  \min \left( \lfloor B_{\rm active} T / 32 \rfloor, W_{\rm max}  \right)
\end{equation}
and $B_{\rm active}$ is given as a flow chart
by Figure~\ref{fig:fig:occupancysimpleflowchart}.
\ifPATENT
{\em Plain English}
explanations to clarify
the underlying reasoning in the definition
of hardware occupancy are given by
Figure~\ref{fig:occupancysimpleflowchart}.
\fi
Figure~\ref{fig:fig:occupancysimpleflowchart} shows
how one can derive a rational program computing
$B_{\rm active}$ from $R_{\rm max}$, $Z_{\rm max}$,
$T_{\rm max}$, $B_{\rm max}$, $W_{\rm max}$, $R$, $Z$, $T$.
It follows from Equation (\ref{eq:occupancy})
that $W_{\rm active}$ can also be computed 
by a rational program from $R_{\rm max}$, $Z_{\rm max}$,
$T_{\rm max}$, $B_{\rm max}$, $W_{\rm max}$, $R$, $Z$, $T$.
Finally, the same is true for the hardware occupancy
of a {\cuda} kernel.

}
\end{example}

\begin{example}
\label{ex:mwpcwp}
{\em 
The {\em execution time} of a GPU kernel is defined 
in the MWP-CWP execution model, 
see \cite{DBLP:conf/isca/HongK09,DBLP:conf/ppopp/SimDKV12}, 
and calculated from hardware parameters including:
\begin{itemizeshort}
\item[-] clock frequency of a SM,
\item[-] the number of SMs on the device, etc.,
\end{itemizeshort}
as well as low-level performance metrics, including:
\begin{itemizeshort}
\item[-] the total number $\#{\rm Mem\_insts}$
of memory instructions per thread,
\item[-] the total number $\#{\rm Comp\_inst }$ of computation 
instructions per thread, etc.,
\end{itemizeshort}
 and a program parameter, namely the 
number $T$ of threads per thread block.
\ifPATENT
The code in Appendix~\ref{app}, which is written in the 
C programming language, lists from line 246 to line 525 provides a concrete example of a 
rational program. 
\else
We have implemented this model  in the C programming language, see Section~\ref{sec:embodiment}.
\fi
This program computes the execution time (variable {\tt clockcycles})
as a function of the hardware parameters, low-level performance
metrics, and program parameters considered by the MWP-CWP execution
model.
}
\end{example}
\section{Technique Principle}
\label{sec:principle}

\subsection{Key observations}
\label{sect:keyobservation}

Section~\ref{sect:steps} presents a technique for constructing a
rational program ${\cal R}$ in $P_1, \ldots, P_p$, $D_1, \ldots, D_d$
evaluating ${\cal E}$. This technique is based on the
following two observations. The first one is 
a general remark about rational programs
while the second one is specific to rational programs
estimating performance metrics.

\begin{observation}
\label{obs:rationality}
{\em 
Let ${\cal S}$ be a rational program in $X_1, \ldots, X_n$ evaluating $Y$.
Let $s$ be any instruction of ${\cal S}$ other than a branch or an
integer part instruction.  Hence, this instruction can be of the form
$Z = -U$, $Z = U + V$, $Z = U - V$, $Z = U \times V$, where $U$ and
$V$ can be any machine-representable rational numbers.
Let $V_1, \ldots, V_v$ be the variables that are defined
at the entry point of the basic block of  the instruction $s$.
An elementary proof by induction yields the following fact.
There exists a rational function\footnote{Here, rational function is in the
  sense of algebra, see~\cite{eisenbud2013commutative} and section \ref{subsec:paramest} below.} in $V_1,
\ldots, V_v$ that we denote by $f_s(V_1, \ldots, V_v)$
such that $Z = f_s(V_1, \ldots, V_v)$ for all possible values of
$V_1, \ldots, V_v$.

From there, one derives the following observation.
There exists a partition ${\cal T} = \{ T_1, T_2, \ldots    \}$ 
of ${\Q}^n$ (where {\Q} denotes the field of rational numbers)
and rational functions $f_1(X_1, \ldots, X_n)$, 
 $f_2(X_1, \ldots, X_n)$, \ldots 
such that, if $X_1, \ldots, X_n$ receive respectively 
the values $x_1, \ldots, x_n$, then
the value of $Y$ returned by ${\cal S}$ is one of
$f_i(x_1, \ldots, x_n)$ where $i$ is such that
$(x_1, \ldots, x_n) \in T_i$ holds.
In other words, ${\cal S}$ computes $Y$ as a
piece-wise rational function.
Figure~\ref{fig:fig:occupancysimpleflowchart} 
shows that the hardware occupancy of a {\cuda}
kernel is given as a piece-wise rational function
in the variables $R_{\rm max}$, $Z_{\rm max}$,
$T_{\rm max}$, $B_{\rm max}$, $W_{\rm max}$, $R$, $Z$, $T$.
Hence, in this example, we have $n = 8$.
Moreover, Figure~\ref{fig:fig:occupancysimpleflowchart} 
shows that its partition of ${\Q}^n$ contains 5 parts.
}
\end{observation}

\begin{observation}
\label{obs:faisability}
{\em 
It is natural to assume that low-level performance metrics 
depend rationally on program parameters and data parameters.
While this latter fact is not explicitly discussed
in the {\cuda} and MWP-CWP models, it is an obvious observation
for very similar models which are based on the 
{\em Parallel Random Access Machine}
(PRAM)~\cite{DBLP:journals/siamcomp/StockmeyerV84,Gibbons:1989:MPP:72935.72953},
including PRAM-models tailored to GPU code analysis
such as 
TMM~\cite{ma2014memory} and  MCM~\cite{DBLP:conf/parco/HaqueMX15}.

Given our assumption that 
the high-level performance metric ${\cal E}$
is computable as a rational program depending on hardware parameters,
low-level performance metrics, 
and program parameters, 
we can therefore 
assume that ${\cal E}$ can be expressed
as a rational program of the hardware parameters, 
the program parameters, 
and the data parameters. 
That is, we replace the direct dependency on low-level metrics
with a dependency on the data and program parameters.
Moreover, when applied to the multithreaded program ${\cal P}$ 
to be executed on a
particular multiprocessor, we can assume that ${\cal E}$ can be expressed
as a rational program depending only on program and data parameters. 
}
\end{observation}

\subsection{Consequences}
\label{sec:consequences}

\begin{consequence}
\label{cons:one}
{\em 
It follows from Observation~\ref{obs:faisability} that we can use
the MWP-CWP performance model on top of \cuda\ (see Examples~\ref{ex:cuda}
and \ref{ex:mwpcwp}) to determine a rational program estimating ${\cal
E}$ depending on only program parameters 
and data parameters. 
Recall that building such a rational
program ${\cal R}$ is our goal.  Once ${\cal R}$ is known, it can be
used at runtime (that is, when the program ${\cal P}$ is run
on specific input data) to compute optimal values for the program
paramters $P_1, \ldots, P_p$ .
}
\end{consequence}

\begin{consequence}
\label{cons:two}
{\em 
Observation~\ref{obs:rationality}
suggests an approach for determining ${\cal R}$.
Suppose that a flow chart ${\cal C}$ 
representing ${\cal R}$ is partially known;
to be precise, suppose that the decision nodes
are known (that is, mathematical expressions
defining them are known) while the process nodes
are not. 
From Observation~\ref{obs:rationality},
each process node that includes no integer part
instructions is given by
a series of rational functions.
Determining each of those rational functions
can be achieved by solving 
an \textit{interpolation} or \textit{curve fitting} problem.
Knowing that a function to-be-determined 
is rational allows us to perform \textit{parameter estimation}
techniques (e.g. \textit{the method of least squares}) to define 
the fitting curve and thus the rational function.
}
\end{consequence}

\begin{consequence}
\label{cons:two}
{\em 
In the above consequence, 
one could even consider relaxing the assumption that the
decision nodes are known.
Indeed, each decision node is given by
a series of rational functions.
Hence, those could be determined by
solving curve fitting problems as well.
However, we shall  not discuss this direction further
since this is not needed in the proposed
technique presented in Sections~\ref{sect:steps}
and \ref{sec:embodiment}.
}
\end{consequence}

\subsection{Algorithm}
\label{sect:steps}
In this section, the notations and hypotheses are the same as in
\ifPATENT
Section~\ref{sec:TheIdea}.  
\fi
Sections~\ref{sect:ratprog}, \ref{sect:keyobservation}
and \ref{sec:consequences}.
In particular, we assume that:
\begin{enumerateshort}
\item[$(i)$] ${\cal E}$
is a high-level performance metric for the multithreaded program ${\cal P}$
(e.g. execution time, memory
consumption, and hardware occupancy),
\item[$(ii)$] ${\cal E}$ is given (by a program execution model, e.g. {\cuda}
or MWP-CWP) as a rational program depending on hardware parameters
$H_1, \ldots, H_h$, low-level performance metrics $L_1, \ldots,
L_{\ell}$, and program parameters $P_1, \ldots, P_p$ (see Examples~\ref{ex:cuda} 
and \ref{ex:mwpcwp}),
\item[$(iii)$] the values of the hardware parameters 
are known at compile-time for ${\cal P}$
while the values of the data parameters $D_1, \ldots, D_d$
are known at runtime for ${\cal P}$, 
\item[$(iv)$] the data and program parameters 
take integer values.
\end{enumerateshort}
Extending $(iv)$ we further assume that the possible values of 
the program
parameters $P_1, \ldots, P_p$ belong to a finite set
$F \,\subset\, \mathbb{Z}^p$. That is to say, the possible values of
($P_1, \ldots, P_p$) are a tuple of the form $(\pi_1, \ldots, \pi_p) \in F$,
with each $\pi_i$ being an integer.
Due to the nature of program parameters they are not necessarily all
free variables (i.e. a program parameter may depend on the value of
another program parameter). See Section~\ref{sec:embodiment} for such
an example.

Following Consequence~\ref{cons:one}, we can compute a
rational program ${\cal R}$ estimating ${\cal E}$ as a function of
$D_1, \ldots, D_d$ and $P_1, \ldots, P_p$.  To do so, we shall first 
compute rational functions $g_1(D_1, \ldots, D_d, P_1, \ldots, P_p)$,
\ldots, $g_{\ell}(D_1, \ldots, D_d, P_1, \ldots, P_p)$, estimating
$L_1, \ldots, L_{\ell}$ respectively.
Once ${\cal R}$ has been computed, it can be used 
when ${\cal P}$ is being executed (thus at runtime)
in order to determine optimal values for $P_1, \ldots, P_p$,
on given values of $D_1, \ldots, D_d$.
The entire process is decomposed below into {\bf six} steps:
the first three take place at compile-time while the 
other three are performed at runtime.

\begin{enumerate}
\item \textbf{Data collection}: 
\ifnewversion
In order to determine each of the rational functions 
$$g_1(D_1, \ldots, D_d, P_1, \ldots, P_p),
\ldots, g_{\ell}(D_1, \ldots, D_d, P_1, \ldots, P_p)$$
by solving a curve fitting problem, we proceed as follows:
\begin{enumerateshort}
\item we select a set of points
$K_1, \ldots, K_k$ in the space of the possible
values of $(D_1, \ldots, D_d$, $P_1, \ldots, P_p)$;
\item we run (or emulate) the program ${\cal P}$ 
       at each point $K_1, \ldots, K_k$ and measure the low-level
       performance metrics $L_1, \ldots, L_{\ell}$; and
\item for each $1 \leq i \leq {\ell}$, we record 
       the values $(v_{i,1}, \ldots, v_{i,k})$ 
       measured for $L_i$ at the respective points $K_1, \ldots, K_k$.
\end{enumerateshort}
\fi
\item \textbf{Rational function estimation}: 
\ifnewversion
For each $1 \leq i \leq {\ell}$, we use the values $(v_{i,1}, \ldots,
v_{i,k})$ measured for $L_i$ at the respective points $K_1, \ldots,
K_k$ to estimate the rational function $g_i(D_1, \ldots, D_d,
P_1, \ldots, P_p)$.  We observe that if $v_{i,1}, \ldots, v_{i,k}$
were known exactly (that is, without error) the rational function
$g_i(D_1, \ldots, D_d, P_1, \ldots, P_p)$ could be determined exactly
via rational function interpolation.  However, in practice, the values
$v_{i,1}, \ldots, v_{i,k}$ are likely to be ``noisy''. Hence,
techniques from numerical analysis, like the method of least squares,
must be used instead. Consequently, we compute a rational function
$\hat{g_i}(D_1, \ldots, D_d, P_1, \ldots, P_p)$ which approximates
$g_i(D_1, \ldots, D_d, P_1, \ldots, P_p)$ when evaluated at the points
$K_1, \ldots, K_k$.
\fi
\item \textbf{Code generation}: 
\ifnewversion
In order to generate the rational program ${\cal R}$, we 
proceed as follows:
\begin{enumerateshort}
\item we convert the rational program representing 
${\cal E}$ into code, say in the C programming language, 
essentially encoding the CFG for computing $\cal{E}$;
\item we convert each of 
$\hat{g_i}(D_1, \ldots, D_d, P_1, \ldots, P_p)$,
(for $1 \leq i \leq {\ell}$)
into code as well, more precisely, into sub-routines
evaluating $L_1, \ldots, L_{\ell}$, respectively; and
\item we include those sub-routines
into the code computing ${\cal E}$, which yields
the desired rational program ${\cal R}$.
\end{enumerateshort}
\fi
\item \textbf{Rational program evaluation}: 
\ifnewversion
At this step, the rational program ${\cal R}$ is completely determined
and can be used when the program ${\cal P}$ is executed. At the time
of execution, the data parameters $D_1, \ldots, D_d$ are given
particular values, say ${\delta}_1, \ldots, {\delta}_d$, respectively.
For those specified values of $D_1, \ldots, D_d$ and for
all \fixed{practically meaningful}{should we say practically
meaningful or principled instead?} values of $P_1, \ldots, P_p$ from
the set $F$, we compute an estimate of ${\cal E}$ using ${\cal R}$.
Here, ``practically meaningful'' refers to the fact that the values of the
program parameters $P_1, \ldots, P_p$ are likely to be constrained by
the values of the data parameters $D_1, \ldots, D_d$.  For instance,
if $P_1, P_2$ are the two dimension sizes of a two-dimensional
thread-block of a {\cuda} kernel performing the transposition of a
square matrix of order $D_1$, then the inequality $P_1 P_2 \leq D_1^2$
is meaningful.  Despite of this remark, this step can still be seen as
an exhaustive search, and, it is practically feasible to do so for
three reasons:
\begin{enumerateshort}
\item $p$ is small, typically $p=1$, $p=2$ or $p=3$,
      see Section~\ref{sec:embodiment};
\item the set of values in $F$ are small 
	(typically on the order of 10 elements in case of MWP-CWP model); and
\item the program ${\cal R}$ simply evaluates 
       a few formulas and thus runs almost instantaneously.
\end{enumerateshort}
\fi
\item \textbf{Selection of optimal values of program parameters}: 
\ifnewversion
In general, when the search space of values of the program parameters
 $P_1, \ldots, P_p$ is large, a numerical optimization technique is required
 for this step. For CUDA kernels, however, selecting a value point
  (also called a {\em configuration} in the rest of this report) which is optimal 
  can be done with an exhaustive search, as seen in the previous step, 
  and is hence trivial.
The only possible challenge in this case is that several 
configurations may optimize ${\cal E}$. When this happens, 
using a second performance metric can help refine the choice 
of a {\em best configuration}.
\fi
\item \textbf{Program execution}: 
\ifnewversion
Once a best configuration is selected, the
program ${\cal P}$ can finally be executed using this
configuration of program parameters $P_1, \ldots, P_p$
along with with the values ${\delta}_1, \ldots, {\delta}_d$
of $D_1, \ldots, D_d$.

\fi
\end{enumerate}

\section{Implementation}
\label{sec:embodiment}
In the previous sections we gave an overview of our technique for
general multithreaded programs $\cal{P}$. For the embodiment of this
technique, and the resulting experimentation and performance analysis,
we focus on programs for GPU architectures using
the programming model {\cuda}.
These programs interleave 
\begin{inparaenum}[(i)]
  \item serial code which is executed
on the host (the CPU) and, multithreaded
code which is  executed on the device (the GPU).
\end{inparaenum}
The host launches a device code execution by calling a particular type
of function, called a {\em kernel}.
Optimizing a {\cuda} program for better usage of the computing 
resources on a device essentially means optimizing each of its kernels.
Therefore, we apply the ideas of Sections~\ref{sect:keyobservation}
through \ref{sect:steps} to {\cuda} kernels.

In the case of a {\cuda} kernel the data parameters
are a description of the input data size.
In many examples this is a single parameter, say $N$,
describing the size of an array (or the order of a two-dimensional array), 
the values of which are usually powers of $2$.
On the other hand, the program parameters are typically an encoding of
the program configuration, namely the \textit{thread block
configuration}. In {\cuda} the thread block configuration defines both
the number of dimensions (1, 2, or 3) and the size in each dimension
of each \textit{thread block} --- a contiguous set of threads mapped
together to a single SM processor. For example, a possible thread block configuration
may be $1024 \times 1 \times 1$ (a one-dimensional thread block),
or $16 \times 16 \times 2$ (a three-dimensional thread block).
In all cases the {\cuda} model restricts the number of threads per block
(i.e. the product of the size in each dimension)
to be $1024$ on devices of Compute Capability (CC) 2.x, or newer. 
This limits the possible thread block configurations, and, moreover, 
limits the set $F$ from which the program parameters 
are chosen.

Throughout the current and the following
sections our discussions make use of these specialized terms, input data size and thread block configuration, 
for data and program parameters, respectively, in order to make clear explanations and associations
between theory and practice.

\subsection{Data collection}
\label{sec:data_collection}
As detailed in Step 1 of Section~\ref{sect:steps}, we must collect
data and statistics regarding certain performance counters and runtime
metrics (which are thoroughly defined in \cite{DBLP:conf/isca/HongK09}
and \cite{cuda2015}). These metrics together allow us to estimate the
execution time of the program and 
can be partitioned into three categories.

{\em First}, we need architecture-specific performance counters,
basically, characteristics dictated by the CC
of the target device. Such hardware characteristics can be obtained
at compile time, as the target CC is specified at this time.
These characteristics include the number of registers used in each thread,
the amount of static shared memory 
per thread block, and the number of (arithmetic, memory, and synchronization)
instructions per thread. This information can easily be obtained from 
a {\cuda} compiler (e.g. NVIDIA's \texttt{NVCC}).

{\em Second}, we have the values that depend on the behavior of the
kernel code at runtime which we will refer to as kernel-specific runtime values.
This includes values impacted by memory access patterns, 
namely, 
the number of coalesced and non-coalesced memory accesses per warp,
the number of memory instructions in each thread that cause
coalesced and non-coalesced accesses, 
and eventually, the total number of warps that are being executed.
For this step we have two choices. 
We can use an emulator, 
which can mimic the behavior of a GPU on a typical Central Processing Unit (CPU). 
This is important if we cannot guarantee that a device
is available at compile time.
For this purpose, we have configured Ocelot \cite{DBLP:conf/IEEEpact/DiamosKYC10},
\ifICPP
a dynamic compilation framework for GPU computing as well as an emulator
for low-level {\cuda} (PTX \cite{cudatoolkit2018}) code,
\fi
to meet our needs.
\ifPATENT
Ocelot is a dynamic compilation framework for GPU computing
which translates from PTX (NVIDIA's low-level parallel thread 
instruction set architecture \cite{cudatoolkit2018}) to
various GPU architectures and multi-core CPU architectures. Ocelot also
provides an emulator for such PTX code. 
Again, we use the \texttt{NVCC} compiler to compile
{\cuda} code to such low-level intermediate PTX code.
\fi
Alternatively, we can use a profiler to collect the required metrics
on an actual device. For this solution, we have used 
NVIDIA's {\nvprof}\cite{nvprofguide}. 

{\em Third}, in order to compute a more precise estimate of the clock cycles,
we need device-specific parameters which describe an actual GPU card.
One subset of such parameters can be determined by microbenchmarking 
the device (see \cite{DBLP:journals/tpds/MeiC17} and \cite{DBLP:conf/ispass/WongPSM10}),
this includes the memory latency, the memory bandwidth, 
and the departure delay for coalesced/non-coalesced accesses.
Another subset of such parameters can be easily obtained
by consulting the vendor's guide \cite{cuda2019guide}, 
or by querying the device itself via the CUDA API.
This includes the number of SMs on the card, 
the clock frequency of SM cores, and the instruction delay. 

In summary, our general method proceeds as follows:
\begin{enumerateshort}
	\item[(1)] Beginning with a {\cuda} program, we minimally annotate the host code
	to make it compatible with our \textit{pre-processor program}, specifying the code fragment
	in the host code which calls the kernel. 
	We also specify the CC of the target device.
	

	\item[(2)] The pre-processor program prepares the code for 
	collecting kernel-specific runtime values.
	This step includes source code analysis in order to extract
	the list of kernels, the list of kernel calls in the host code, 
	and finally, the body of each kernel (which will be used for further analysis).
	In the case of Ocelot emulation, 
	the pre-processor replaces the device kernel calls with 
	equivalent Ocelot kernel calls, while in the case of actual profiling,
	the device kernel calls are left untouched.
	Finally, the pre-processor program uses the specified CC 
	in the host code to determine the architecture-specific 
	performance counters for each kernel.
	The result of this step is an executable file, which we will refer to as
	\textit{the instrumentor}. The instrumentor takes as input the same program
    parameters as the original {\cuda} code.

	\item[(3)] A driver program orchestrates the combination of 
	device-specific characteristics (i.e.\ a device profile) and various
	configurations of program and data parameters to be passed to the instrumentor.
	Running the instrumentor (either emulated on top of Ocelot, 
	or running via a profiler on a GPU) 
	measures and records the required performance metrics.
	
\end{enumerateshort}

\subsection{Rational function estimation}
\label{subsec:paramest}
Once data collection is completed, we are able to move
on to the estimation step, see Step 2 of Section~\ref{sect:steps}.
That is, determining precisely the rational functions
$f_b(X_1,\dots,X_n)$, for each desired code block $b$, which are to be
used within the rational program (see Section~\ref{sect:ratprog}).
These $X_i$ variables are simply combinations of program parameters,
data parameters, and intermediate values, obtained during the execution
of the rational program.
These intermediate values are, in turn, are rational functions of 
the program and data parameters. Hence, the variables $X_1, \dots, X_n$ are
used only to simplify notations.
\begin{align*}
f_b(X_1,\dots,X_n) &= \frac{p_b(X_1,\dots,X_n)}{q_b(X_1,\dots,X_n)} \\
&= \frac{\alpha_1\cdot(X_1^0\dots X_n^0) \;+\; 
	\ifPATENT \alpha_2\cdot(X_1^1\dots X_n^0) \;+\; \fi
	\dots \;+\;  \alpha_i\cdot(X_1^{u_1}\dots X_n^{u_n})}{\beta_1\cdot(X_1^0\dots X_n^0) \;+\;
	\ifPATENT  \beta_2\cdot(X_1^1\dots X_n^0) \;+\; \fi
	\dots \;+\; \beta_j\cdot(X_1^{v_1}\dots X_n^{v_n})}
\end{align*}

A rational function is simply a fraction of two polynomials. With
a \textit{degree bound} (an upper limit on the exponent) on each
variable $X_k$ in the numerator and the denominator,
say $u_k$ and $v_k$, respectively, these polynomials can be defined up to
some \textit{parameters}, namely the coefficients of the polynomials,
say $\alpha_1,\dots,\alpha_i$ and $\beta_1,\dots,\beta_j$.

Using the previously collected data (see Section~\ref{sec:data_collection})
we perform a parameter estimation (for each rational function)
on the polynomial coefficients $\alpha_1, \ldots, \alpha_i, \beta_1, \ldots, \beta_j$. 
in order to uniquely determine the rational function. 
We note that while our model is non-linear, it is indeed linear in 
the model-fitting parameters $\alpha_1, \ldots, \alpha_i, \beta_1, \ldots, \beta_j$. 

The system of linear equations defined by 
the model-fitting parameters $\alpha_1, \ldots, \alpha_i, \beta_1, \ldots, \beta_j$
can very classically be defined as an equation
of the form  $\mathbf{Ax} = \mathbf{b}$,
where $\mathbf{A} \in \mathbb{R}^{m\times n}$ (the \textit{sample matrix}) and 
$\mathbf{b} \in \mathbb{R}^{m \times 1}$ (the right-hand side vector)
encode the collected data, while the solution vector 
$\mathbf{x} \in \mathbb{R}^{n \times 1}$ encodes the 
model-fitting parameters\footnote{Keen observers will notice that, for rational functions,
we must actually solve a system of homogeneous equations.
Such details are omitted here, but we refer the reader to \cite[Chapter 5]{brandt2018high}.}.
An exact solution is rarely defined if $m < n$ (where an infinite number of solutions is 
possible) or if $m > n$. Therefore, we wish to 
get a solution in the ``least squares sense'', that is, find 
$\mathbf{x}$ such that \textit{residual} is minimized: 
\begin{gather*}
\mathbf{x} \:=\: \min_{\mathbf{x}}||\mathbf{r}||^2_2 \:=\: \min_{\mathbf{x}}||\mathbf{b} - \mathbf{Ax}||_2^2
\end{gather*} 

Many different methods exist for solving this so-called linear least squares problem, 
such as the \textit{normal-equations}, or \textit{QR-factorization}, 
however, these methods are either numerically unstable (normal-equations), or will fail
if the sample matrix is rank-deficient (both normal-equations and QR) 
\cite{corless2013graduate}.
We rely then on the \textit{singular value decomposition} (SVD) of $\mathbf{A}$ 
to solve this problem.
This decomposition is very computationally intensive, much more than that of
normal-equations or QR, but is also much more numerically
stable, as wel as being capable of producing solutions with a rank-deficient sample matrix.

We are highly concerned with the robustness of our
method due to three problems present in our particular situation:
\begin{enumerate}
	\item[(1)] the sample matrix is very ill-conditioned;
	\item[(2)] the sample matrix will often be (numerically) rank-deficient;
	\item[(3)] we are interested in using our fitted model for extrapolation, meaning 
	any numerical error in the model fitting will grow very quickly \cite{corless2013graduate}.
\end{enumerate}
While (3) is an issue inherent to our model fitting problem, (1) and (2) result
from our choice of model, and how the sample points $(X_1,\ldots, X_n)$ are chosen, respectively.
Using a rational function (or polynomial) as the model for which we wish to estimate parameters
presents numerical problems. The resulting sample matrix is essentially 
a Vandermonde matrix. These matrices,
while theoretically of full rank, are extremely ill-conditioned
\cite{corless2013graduate, beckermann2000condition}.
To discuss (2) we must now consider the so-called ``poised-ness'' of the sample points.

For a univariate polynomial model, a sample matrix is 
guaranteed to have full rank if the sample points are all distinct.
We say that the points producing the 
sample matrix are \textit{poised} is the matrix is of full rank.
In the multivariate case,
it is much more difficult to make such a guarantee \cite{chung1977lattices, olver2006multivariate}.
Given a multivariate polynomial model with 
a \textit{total} degree $d$,
a set of points is poised
if, for each point $x_i$, there exists $d$
hyperplanes corresponding to this point such that all other points 
lie on at least one of these hyperplanes but $x_i$ does not \cite{chung1977lattices}. 
For example, with 2 variables and a total degree 3, the sample points are poised
if, for each sample point, there exists 3 lines on which all other points lie but the
selected point does not.

This geometric restriction on sample points is troublesome for our purposes. 
For example, if we are trying to model a function which has, in part,
\cuda\ thread block dimensions as variables, then these dimensions should be
chosen such that their product is a multiple of 32 \cite{ryoo2008optimization}.
However, this corresponds very closely with the geometric restriction we
are trying to avoid.
Consider the possible 2D thread block dimensions ($B_0, B_1$)
for a \cuda\ kernel (Figure~\ref{fig:meshPointsPlot}).
Many subsets of these points are collinear,
presenting challenges for obtaining poised sample points.
As a result of all of this we are highly likely to obtain a rank-deficient
sample matrix.

\input{mesh_points.tex}

Despite all of these challenges our parameter estimation techniques are well-implemented
in optimized C code. We use optimized algorithms from LAPACK (Linear Algebra PACKage) \cite{userguide:lapack}
for singular value decomposition and linear least squares solving
while rational function and polynomial implementations are similarly highly optimized
thanks to the Basic Polynomial Algebra Subprograms (BPAS) library \cite{bpasweb, brandt2018high}.
With parameter estimation complete the rational functions
required for the rational program are fully specified
and we can finally construct it.

\subsection{Rational programs}

In practice, the use of rational programs is split into two parts:
the generation of the rational program at the compile-time of the multithreaded 
program $\cal{P}$, 
and the use of the rational program during the runtime of $\cal{P}$.

\subsubsection{Compile-time code generation}
We are now at Step 3 of Section~\ref{sect:steps}.  We look to define a
rational program which evaluates the high-level metric $\cal{E}$ of the
program $\cal{P}$ using the \mwpcwp\ model.  In implementation, this
is achieved by using a previously defined \textit{rational program
template} which contains the formulas and case discussion of
the \mwpcwp\ model, independent of the particular program being
investigated. Using simple regular expression matching and text
manipulation we combine the rational program template with the rational functions 
determined in the previous step to obtain a rational program
specialized to the multithreaded program $\cal{P}$. The generation of
this rational program is performed completely during compile-time,
before the execution of the program itself.

\subsubsection{Runtime optimization}
At runtime, the input data sizes (data parameters) are well known.
In combination with the known hardware parameters, since the program
is actually being executed on a specific device,
we are able to specialize every parameter in the rational program
and obtain an estimate for the high-level metric $\cal{E}$.
This rational program is then easily and quickly evaluated during
(or immediately prior to) the execution of $\cal{P}$. Evaluating 
the rational program for each possible thread block configuration,
as restricted by our data parameters
and the \cuda\ programming model itself, we determine a thread block configuration
which optimizes $\cal{E}$. The program $\cal{P}$ is finally executed 
using this optimal thread block configuration. 
Therefore, we have completed Steps 4 to 6 of Section~\ref{sect:steps}.



\section{Experimentation}
\label{sec:performance}
In this section we highlight the performance and experimentation of
our technique applied to example {\cuda} programs
from the PolyBench benchmarking suite \cite{polybenchGPUweb}. 
For each PolyBench example,
we collect statistical data (see Section~\ref{sec:embodiment}) with
various input sizes $N$ (powers of 2 typically between 64 and 512). 
These are relatively small sizes to allow for fast data collection.
The collected data is used 
to produce the rational functions and, lastly, the rational functions
are combined to form the rational programs.
Table~\ref{tab:example_names} gives a short description for
each of the benchmark examples, as well as an index for each kernel
which we use for referring to that kernel in the experimental results.

The ability of the rational programs we generate to effectively
optimize the program parameters of each example {\cuda} program is
summarized in Tables~\ref{tab:sanitycheck_nvprof}
and \ref{tab:proofofconcept_nvprof}. The definitions of the notations used
in these tables are as follows:
\begin{itemizeshort}
	\item $N$: the input size of the problem, usually a power of 2,
	\item $Ec$: estimated execution time of the kernel measured in clock-cycles,
	\item $C_r$: best thread block configuration as decided by the {\em rational program},
	\item $C_d$: default thread block configuration given by the benchmark,
	\item $Ec_r$: $Ec$ of configuration $C_r$ as decided by the {\em rational program},
	\item $C_i$: best thread block configuration as decided by \textit{instrumentation},
	\item $Ec_i$: $Ec$ of configuration $C_i$ as decided by \textit{instrumentation},
	\item {\rm Best\_time (B\_t)}: the best {\cuda} running time 
	(i.e. the time to actually run the code on a GPU) of the kernel among all possible configurations,
	\item {\rm Worst\_time (W\_t)}: the worst {\cuda} running time of the kernel among all possible configurations.
\end{itemizeshort}

To evaluate the effectiveness of our rational program, we compare 
the results obtained from it with an actual analysis of the program 
using the performance metrics collected by the instrumentor. 
This comparison gives us estimated clock cycles as shown in 
Table~\ref{tab:sanitycheck_nvprof}.
The table shows the best configurations ($C_i$ and
$C_r$) for $N = 256$ and $N = 512$ for each example, as decided from
both the data collected during instrumentation and the resulting rational
program.  This table is meant to highlight that the construction of
the rational program from the instrumentation data is performed
correctly. Here, either the configurations should be the same or, if
they differ, then the estimates of the clock-cycles should be very
similar; indeed, it is possible that different thread block
configurations (program parameters) result in the same execution
time. Hence, the rational program must choose one of the possibly many
optimal configurations. Moreover, we supply the value ``Collected
$Ec$'' which gives the $Ec$ given by the instrumentation for the thread
block configuration $C_r$.  This value shows that the rational program
is correctly calculating $Ec$ for a given thread block
configuration. The values ``Collected $Ec$'' and $Ec_r$ should be very
close.

Figure~\ref{fig:atax_kernel2-512}
illustrates detailed results for Kernel 8.2 when $N = 512$. 
The ``\cuda'' plot shows the real runtime of the kernel launched
on a device while the ``RP-Estimated'' plot shows the estimates
obtained from our rational program.
The $x$ axis shows the various thread block configurations.
The runtime values on $y$ axis are normalized to values between 
$0$ and $100$.  

Table~\ref{tab:proofofconcept_nvprof} shows the best configuration ($C_r$)
for $N = 1024$ and $N = 2048$ estimated by the rational program of
each example.  Notice these values of $N$ are much larger than those
used during the instrumentation and construction of the rational
program. This shows that the rational program can extrapolate on the
data used for its construction and accurately estimate optimal program
parameters for different data parameters.  To compare the thread block
configuration deemed optimal by the rational program ($C_r$) and the
optimal thread block configuration determined by actually executing
the program with various thread block configurations we calculate the
value named ``Error''.  This is a  percentage  given
by ($C_rt$ - Best\_time)/(Worst\_time - Best\_time) *
100\%, where ``$C_rt$'' is the actual running time of the program
using the thread block configuration $C_r$ in millisecond. Best\_time  and Worst\_time
are as defined above.  The error percentage shows the
quality of our estimation; it clearly shows the difference between the
estimated configuration and the actual best configuration.  This
percentage should be as small as possible to indicate a good
estimation. 

As an exercise in the practicality of our technique,
 we also show the default thread block configuration ($C_d$) 
 given by the benchmarks  and ``$C_dt$'', the actual running time of the program
using the thread block configuration $C_d$ in the same table. 
Compared to the default configurations, the rational program determines a better configuration
for more than half of the kernels.
Notice that, in most cases, the same default configuration is used 
for all kernels within a single example.
In contrast, our rational program can produce different configurations optimized
for each kernel individually.

As mentioned in Section~\ref{sec:data_collection}, we use
both Ocelot \cite{DBLP:conf/IEEEpact/DiamosKYC10} and
NVIDIA's {\nvprof}\cite{nvprofguide} as instrumentors in the data collection step.
The ``Error''  of the implementation based on data collected by Ocelot is shown in the column ``Ocel'' in Table~\ref{tab:proofofconcept_nvprof}.
We can see that the overall performance is better with \nvprof,
but for some benchmarks, for example Kernel 12,
the best configuration picked by the rational program 
is not as expected. We attribute this to the limitations of the
underlying MWP-CWP model we use.

More experimental results may be found on our GitHub repository:
{
\small
\color{navy}
\url{https://github.com/orcca-uwo/Parametric_Estimation/tree/master/plots}
}

\ifPATENT
\afterpage{
	\thispagestyle{empty}
\fi
	\begin{table}[htb]
		\centering
		\ifPATENT \small \fi
                \ifICPP \scriptsize \fi
		\begin{tabular}{|l|l|l|l|}
			\hline
			Benchmark & Description & Kernel name & Kernel ID\\
			\hline
			2D Convolution & 2-D convolution&Convolution2D\_kernel&$1$\\
			\hline
			\multirow{3}{*}{FDTD\_2D}&{2-D Finite Different}&fdtd\_step1\_kernel & $2.1$\\ \cline{3-4}
			&\multirow{2}{*}{Time Domain}&{fdtd\_step2\_kernel}& $2.2$	\\ \cline{3-4}
			&&{fdtd\_step3\_kernel} & $2.3$ \\ 
			\hline
			2MM & 2 Matrix Multiplications& mm2\_kernel1 &$3$\\
			\hline
			3MM & 3 Matrix Multiplications & mm3\_kernel1& $4$\\
			\hline
			\multirow{2}{*}{BICG}&\multirow{2}{*}{BiCGStab Linear Solver}&bicg\_kernel1 & $5.1$\\ \cline{3-4}
			&&bicg\_kernel2& $5.2$	\\ 
			\hline
			GEMM &Matrix Multiplication&gemm\_kernel& $6$\\
			\hline
			3D\_Convolution&3-D Convolution&convolution3D\_kernel & $7$\\
			\hline
			\multirow{2}{*}{ATAX}&Matrix Transpose and &atax\_kernel1 & $8.1$\\ \cline{3-4}
			&Vector Multiplication&atax\_kernel2& $8.2$	\\ 
			\hline
			
			\multirow{2}{*}{GESUMMV} & Scalar, Vector and &\multirow{2}{*}{gesummv\_kernel}&\multirow{2}{*}{$9$}\\
			&Matrix Multiplication&&\\
			\hline
			{SYRK} & Symmetric rank-k operations&syrk\_kernel&$10$\\
			\hline
			\multirow{2}{*}{MVT}&{Matrix Vector Product and}&mvt\_kernel1 & $11.1$\\ \cline{3-4}
			&Transpose&mvt\_kernel2& $11.2$	\\ 
			\hline
			SYR2K & Symmetric rank-2k operations&syr2k\_kernel& $12$\\
			\hline
			\multirow{4}{*}{CORR}&\multirow{4}{*}{Correlation Computation}&corr\_kernel & $13.1$\\ \cline{3-4}
			&&mean\_kernel & $13.2$\\ \cline{3-4}
			&&reduce\_kernel & $13.3$\\ \cline{3-4}
			&&std\_kernel& $13.4$	\\ 
			\hline
			\multirow{3}{*}{COVAR}&\multirow{3}{*}{Covariance Computation}&covar\_kernel & $14.1$\\ \cline{3-4}
			&&mean\_kernel & $14.2$\\ \cline{3-4}
			&&reduce\_kernel & $14.3$\\ 
			\hline
			\multirow{3}{*}{GRAMSCHM}&\multirow{3}{*}{Gram-Schmidt decomposition}&gramschmidt\_kernel1 & $15.1$\\ \cline{3-4}
			&&gramschmidt\_kernel2 & $15.2$\\ \cline{3-4}
			&&gramschmidt\_kernel3 & $15.3$\\ 
			\hline
		\end{tabular}
		\caption{\rm Benchmark names and descriptions}
		\label{tab:example_names}
	\end{table}
\ifPATENT
	\clearpage
}
\fi




\input{atax_512}

\ifPATENT
\afterpage{
\thispagestyle{empty}
\fi
\begin{table}[htb]
	\centering
		\ifPATENT \small \fi
                \ifICPP \scriptsize \fi
	\begin{tabular}{|l|l|l|l|l|l|l|}
		\hline
		{\multirow{2}{*}{ID}}
		 & \multirow{2}{*}{$N$} & \multicolumn{2}{l|}{Data collection} &  \multicolumn{3}{l|}{Rational Program} \\
		\cline{3-7}
		& &$C_i$&$Ec_i$& $C_r$ & $Ec_r$ & Collected $Ec$ \\
		\hline
		\multirow{2}{*}{1}
		& 256 & $16 \times 2$& 198070 & $16 \times 2$ & 310463 & 198070 \\
		& 512 & $16 \times 4$& 794084 & $16 \times 16$ & 947799 & 889378 \\ \hline
		\multirow{2}{*}{2.1}  
		& 256 & $16 \times 2$& 44336 & $32 \times 1$ & 40213 & 44336 \\
		& 512 & $16 \times 4$& 147622 & $32 \times 2$ & 135821 & 147622 \\
		\cline{2-7}
		\multirow{2}{*}{2.2}  
		& 256 & $32 \times 1$& 19430 & $32 \times 1$ & 52096 & 19430 \\
		& 512 & $32 \times 2$& 64475 & $64 \times 1$ & 135810 & 64475 \\
		\cline{2-7}
		\multirow{2}{*}{2.3}  
		& 256 & $32 \times 1$& 19476 & $16 \times 2$ & 97964 & 19640 \\
		& 512 & $32 \times 2$& 64643 & $16 \times 4$ & 326474 & 65191 \\
		\hline
		{\multirow{2}{*}{3}} 
		& 256 & $16 \times 2$& 351845192 & $32 \times 1$ & 1365006816 & 351845192 \\
		& 512 & $16 \times 4$& 4610508174 & $16 \times 64$ & 18148896630 & 4610508174 \\
		\hline
		{\multirow{2}{*}{4}} 
		& 256 & $16 \times 2$& 351845192 & $16 \times 2$ & 1364711040 & 351845192 \\
		&512 & $16 \times 4$& 4610508174 & $16 \times 64$ & 18147912950 & 4610508174 \\
		\hline
		\multirow{2}{*}{5.1}
		& 256 & $8 \times 4$& 6730198 & $16 \times 2$ & 13105400 & 6927829 \\
		& 512 & $16 \times 2$& 27224663 & $32 \times 1$ & 56699908 & 28801623 \\
		\cline{2-7}
		\multirow{2}{*}{5.2} 
		& 256 & $8 \times 4$& 6729688 & $16 \times 2$ & 13105400 & 6927319 \\
		& 512 & $16 \times 2$& 27223641 & $32 \times 1$ & 56699908 & 28800601 \\  
		\hline
		{\multirow{2}{*}{6}} 
	& 256 & $16 \times 2$& 352450569 & $32 \times 1$ & 1365746112 & 352450569 \\
	& 512 & $16 \times 4$& 4614492176 & $16 \times 4$ & 18091801600 & 4614492176 \\ 
		\hline
		{\multirow{2}{*}{7}} 
& 256 & $16 \times 2$& 412448 & $8 \times 4$ & 419355 & 423967 \\
& 512 & $16 \times 4$& 1382453 & $16 \times 4$ & 1382446 & 1382453 \\ 
		\hline
		\multirow{2}{*}{8.1}
& 256 & $8 \times 4$& 6741283 & $16 \times 2$ & 13836355 & 6938145 \\
& 512 & $16 \times 2$& 27245283 & $32 \times 1$ & 57486243 & 28819169 \\
		\cline{2-7}
		\multirow{2}{*}{8.2} 
& 256 & $8 \times 4$& 6741791 & $16 \times 2$ & 13836672 & 6938653 \\
& 512 & $16 \times 2$& 27246303 & $32 \times 1$ & 57486880 & 28820189 \\
		\hline
		{\multirow{2}{*}{9}} 
& 256 & $8 \times 4$& 52142036 & $16 \times 2$ & 104085531 & 53194198 \\
& 512 & $16 \times 2$& 211244951 & $32 \times 1$ & 438726895 & 219647900 \\
		\hline
		{\multirow{2}{*}{10}} 
& 256 & $16 \times 2$& 352450567 & $32 \times 1$ & 1365746112 & 352450567 \\
& 512 & $16 \times 4$& 4614492172 & $16 \times 4$ & 18091801600 & 4614492172 \\ 
		\hline
		\multirow{2}{*}{11.1}
& 256 & $8 \times 4$& 6741790 & $16 \times 2$ & 13450360 & 6938652 \\
& 512 & $16 \times 2$& 27246302 & $32 \times 1$ & 56715288 & 28820188 \\
		\cline{2-7}
		\multirow{2}{*}{11.2} 
& 256 & $8 \times 4$& 6741283 & $16 \times 2$ & 13450360 & 6938145 \\
& 512 & $16 \times 2$& 27245283 & $32 \times 1$ & 56715288 & 28819169 \\
		\hline
		{\multirow{2}{*}{12}} 
& 256 & $16 \times 2$& 1090910216 & $16 \times 2$ & 4293143808 & 1090910216 \\
& 512 & $16 \times 4$& 14402949134 & $32 \times 16$ & 57121538780 & 14402949134 \\
		\hline
		\multirow{2}{*}{13.1}
& 256 & $8 \times 4$& 29762 & $16 \times 2$ & 431621481191 & 34578 \\
& 512 & $1 \times 32$& 81924 & $32 \times 1$ & 7075540802475 & 226841 \\
		\cline{2-7}
		\multirow{2}{*}{13.2}
& 256 & $8 \times 4$& 159563703552 & $16 \times 2$ & 5176052 & 222555249312 \\
& 512 & $16 \times 2$& 3428474954532 & $32 \times 1$ & 22396088 & 3534048733179 \\ 
		\cline{2-7}
		\multirow{2}{*}{13.3}
& 256 & $8 \times 4$& 6938660 & $1 \times 32$ & 5355 & 33687503 \\
& 512 & $16 \times 2$& 28036520 & $1 \times 64$ & 22447 & 502559959 \\
		\cline{2-7}
		\multirow{2}{*}{13.4}
		& 256 & $8 \times 4$& 2526616 & $16 \times 2$ & 14040368 & 2659225 \\
& 512 & $16 \times 2$& 10298524 & $32 \times 1$ & 58770372 & 11353249 \\
		\hline
		\multirow{2}{*}{14.1}  
& 256 & $8 \times 4$& 2526616 & $32 \times 1$ & 474507050858 & 2924449 \\
& 512 & $16 \times 2$& 10298524 & $32 \times 1$ & 7115480129956 & 11353249 \\
		\cline{2-7}
		\multirow{2}{*}{14.2}  
& 256 & $8 \times 4$& 162030833356 & $16 \times 2$ & 5147212 & 225920838132 \\
& 512 & $16 \times 2$& 3454820728440 & $32 \times 1$ & 22398140 & 3560802413900 \\
		\cline{2-7}
		\multirow{2}{*}{14.3}  
& 256 & $64 \times 1$& 1060 & $32 \times 1$ & 1484 & 1069 \\
& 512 & $128 \times 1$& 1085 & $128 \times 1$ & 1677 & 1085 \\
		\hline
		\multirow{2}{*}{15.1}  
& 256 & $16 \times 2$& 77754074 & $256 \times 1$ & 70701 & 646426704 \\
& 512 & $32 \times 1$& 324346910 & $512 \times 1$ & 136765 & 5169397920 \\
		\cline{2-7}
		\multirow{2}{*}{15.2}  
& 256 & $256 \times 1$& 68119 & $8 \times 4$ & 1871 & 1145070 \\
& 512 & $512 \times 1$& 136231 & $8 \times 8$ & 1873 & 8409096 \\
		\cline{2-7}
		\multirow{2}{*}{15.3}  
& 256 & $16 \times 4$& 1492 & $16 \times 2$ & 78089365 & 1505 \\
& 512 & $32 \times 2$& 1492 & $32 \times 1$ & 324813912 & 1505 \\
		\hline			
	\end{tabular}
	\caption{\rm Sanity check (nvprof)}
	\label{tab:sanitycheck_nvprof}
\end{table}
\ifPATENT
\clearpage
}
\fi
\ifPATENT
\afterpage{
\thispagestyle{empty}
\fi
\begin{table}[htb]
	\centering
		\ifPATENT \small \fi
                \ifICPP \scriptsize \fi
	\begin{tabular}{|l|l|l|l|l|l|l|l|l|l|}
		\hline
		{\multirow{2}{*}{ID}}
		& $N$ & \multicolumn{2}{c|}{Default Config}&\multicolumn{2}{c|}{Rational Program} &{\multirow{2}{*}{B\_t}}  &{\multirow{2}{*}{W\_t}}  & 
{\multirow{2}{*}{Error}} &{\multirow{2}{*}{Ocel}}\\
		\cline{3-6}
		&&$C_d$ & $C_dt$& $C_r$ & $C_rt$ &&&&\\
		\hline
		{\multirow{2}{*}{1}} 
		& 1024 & $32 \times 8$& 0.04& $32 \times 8$ & 0.04 & 0.04& 0.34& 0.00 &6.11 \\
		& 2048 & $32 \times 8$&0.16& $128 \times 4$ & 0.16 & 0.15& 1.31& 0.86 &6.43\\
		\hline
		\multirow{1}{*}{2.1}
		& 1024 & $32 \times 8$&18.2& $32 \times 1$ & 23.4 & 18.1& 84.0& 7.93 &0.06\\
		\cline{2-10}
		\multirow{1}{*}{2.2}  
		& 1024& $32 \times 8$& 19.9 & $64 \times 2$ & 19.7 & 19.5& 82.3& 0.26 &0.04\\
		\cline{2-10}
		\multirow{1}{*}{2.3}  
		& 1024& $32 \times 8$& 25.7 & $16 \times 2$ & 27.2 & 25.7& 109& 1.79 &0.02\\
		\hline
		{\multirow{2}{*}{3}}   
		& 1024 & $32 \times 8$&5.73& $16 \times 2$ & 9.67 & 5.73& 89.0& 4.73 &0.57\\
		& 2048& $32 \times 8$& 46.7 & $16 \times 2$ & 78.6 & 46.7& 707& 4.83 &1.10\\
		\hline
		{\multirow{2}{*}{4}}   
		& 1024& $32 \times 8$&5.77 & $16 \times 2$ & 9.68 & 5.73& 89.8& 4.69 &0.63\\
		& 2048 & $32 \times 8$&47.3& $16 \times 2$ & 78.8 & 47.1& 731& 4.63 &1.30\\
		\hline
		\multirow{2}{*}{5.1}
		& 1024 & $256 \times 1$&0.25& $32 \times 1$ & 0.25 & 0.25& 9.27& 0.01 &0.08\\
		& 2048& $256 \times 1$&0.51 & $32 \times 1$ & 0.50 & 0.50& 37.2& 0.00 &1.10\\
		\cline{2-10}
		\multirow{2}{*}{5.2}  
		& 1024 & $256 \times 1$&0.19 & $32 \times 1$ & 0.18 & 0.16& 9.48& 0.17 &0.00\\
		& 2048  & $256 \times 1$& 0.38& $32 \times 1$ & 0.36 & 0.36& 38.5& 0.01 &0.02\\
		\hline
		{\multirow{2}{*}{6}}   
		& 1024 & $32 \times 8$& 5.76& $32 \times 1$ & 9.69 & 5.76& 95.6& 4.37 &0.02\\
		& 2048 & $32 \times 8$&46.3  & $32 \times 1$ & 78.3 & 46.3& 725& 4.70 &1.26\\
		\hline
		{\multirow{1}{*}{7}}   
		& 1024& $32 \times 8$&56.0 & $8 \times 4$ & 81.6 & 55.2& 369& 8.38 &1.49\\
		\hline
		\multirow{2}{*}{8.1} 
		& 1024 & $256 \times 1$& 0.20& $32 \times 1$ & 0.19 & 0.17& 9.87& 0.15 &0.36\\
		& 2048  & $256 \times 1$& 0.39& $32 \times 1$ & 0.37 & 0.37& 42.5& 0.00 &0.28\\
		\cline{2-10}
		\multirow{2}{*}{8.2}  
		& 1024 & $256 \times 1$&0.26  & $32 \times 1$ & 0.24 & 0.24& 9.35& 0.03 &0.42\\
		& 2048& $256 \times 1$&0.50 & $32 \times 1$ & 0.49 & 0.49& 40.3& 0.01 &0.67\\
		\hline
		{\multirow{2}{*}{9}}  
		&1024& $256 \times 1$& 0.40 & $32 \times 1$ & 0.37 & 0.34& 2.10& 1.53 &1.04\\
		& 2048 & $256 \times 1$&0.80  & $32 \times 1$ & 0.74 & 0.73& 6.32& 0.08 &0.44\\
		\hline
		{\multirow{2}{*}{10}}   
& 1024& $32 \times 8$& 0.02 & $32 \times 1$ & 34.3 & 17.9& 87.9& 23.38 &0.00\\
& 2048& $32 \times 8$& 239 & $32 \times 1$ & 968 & 143& 2802& 31.03 &1.39\\
		\hline
		\multirow{2}{*}{11.1}  
& 1024  & $256 \times 1$&0.20& $32 \times 1$ & 0.19 & 0.17& 9.83& 0.17 &0.00\\
& 2048 & $256 \times 1$&0.39  & $32 \times 1$ & 0.38 & 0.36& 39.3& 0.03 &0.03\\
		\cline{2-10}
		\multirow{2}{*}{11.2}  
& 1024 & $256 \times 1$& 0.25 & $32 \times 1$ & 0.24 & 0.24& 9.55& 0.02&0.05 \\
& 2048& $256 \times 1$& 0.50 & $32 \times 1$ & 0.49 & 0.49& 37.4& 0.00 &0.15\\
		\hline
		{\multirow{2}{*}{12}}  
& 1024 & $32 \times 8$&90.5 & $32 \times 1$ & 338 & 32.0& 338& 100.0 &0.25\\
& 2048 & $32 \times 8$&2295& $32 \times 1$ & 5276 & 251& 6723& 77.64 &0.89\\
		\hline
		\multirow{2}{*}{13.1}  
& 1024& $256 \times 1$&320  & $8 \times 8$ & 256 & 224& 2461& 1.42 &0.00\\
& 2048 & $256 \times 1$&1359 & $1 \times 128$ & 7410 & 1186& 18361& 36.23&0.63 \\
		\cline{2-10}
		\multirow{2}{*}{13.2}  
& 1024 & $256 \times 1$&0.26 & $32 \times 1$ & 0.26 & 0.25& 1.98& 0.17 &6.21\\
& 2048& $256 \times 1$& 0.52 & $32 \times 1$ & 0.51 & 0.51& 7.22& 0.00 &1.85\\
		\cline{2-10}
		\multirow{2}{*}{13.3}  
& 1024& $32 \times 8$& 0.05  & $1 \times 128$ & 0.02 & 0.02& 0.21& 0.00 &3.02\\
& 2048& $32 \times 8$& 0.21  & $1 \times 128$ & 0.10 & 0.10& 0.85& 0.13&0.46 \\
		\cline{2-10}
		\multirow{2}{*}{13.4}  
& 1024  & $256 \times 1$&0.26 & $32 \times 1$ & 0.25 & 0.25& 2.18& 0.15 &0.09\\
& 2048  & $256 \times 1$&0.53& $32 \times 1$ & 0.52 & 0.52& 8.16& 0.00 &1.43\\
		\hline
		\multirow{2}{*}{14.1}  
& 1024 & $256 \times 1$& 1065& $32 \times 8$ & 394 & 226& 2462& 7.54 &0.00\\
& 2048& $256 \times 1$& 2116 & $64 \times 16$ & 6119 & 1182& 18343& 28.76 &7.18\\
		\cline{2-10}
		\multirow{2}{*}{14.2}  
& 1024 & $256 \times 1$& 1.63& $64 \times 1$ & 0.26 & 0.25& 2.30& 0.39 &0.11\\
& 2048 & $256 \times 1$& 3.29 & $32 \times 1$ & 0.51 & 0.51& 7.58& 0.05 &1.39\\
		\cline{2-10}
		\multirow{2}{*}{14.3}  
& 1024 & $32 \times 8$&  0.03& $128 \times 1$ & 0.01 & 0.01& 1.48& 0.00 &0.07\\
&2048 & $32 \times 8$&  0.11& $128 \times 1$ & 0.01 & 0.01& 5.94& 0.00 &0.00\\
		\hline
		\multirow{2}{*}{15.1}  
& 1024& $256 \times 1$& 17.3 & $512 \times 1$ & 17.5 & 17.1& 23.1& 7.27 &6.91\\
& 2048& $256 \times 1$& 82.8 & $512 \times 1$ & 83.1 & 82.7& 87.3& 7.98 &24.15\\
		\cline{2-10}
		\multirow{2}{*}{15.2}  
& 1024 & $256 \times 1$&3.03& $64 \times 1$ & 2.93 & 2.93& 4.34& 0.00 &27.98\\
& 2048& $256 \times 1$&6.53  & $32 \times 1$ & 6.64 & 6.36& 11.5& 5.47 &24.56\\
		\cline{2-10}
		\multirow{2}{*}{15.3}  
& 1024& $256 \times 1$& 389 & $32 \times 1$ & 375 & 375& 522& 0.00 &10.42\\
& 2048 & $256 \times 1$&2000  & $32 \times 1$ & 1947 & 1944& 3818& 0.11 &8.63\\
		\hline
	\end{tabular}
	\caption{\rm Proof of Concept (nvprof)}
	\label{tab:proofofconcept_nvprof}
\end{table}


\section{Related work}
\label{sec:relatedwork}
\ifPATENT
\subsection{Research articles in academic venues}
\fi

We briefly review a number of previous works
on the performance estimation of parallel programs
in the context of GPUs and {\cuda} programming model.
Ryoo et al~\cite{RRSBUSH2008} present a performance 
tuning approach for general purpose applications on GPUs 
which is limited to pruning the optimization space of 
the applications and lacks support for memory-bound cases.
Hong et al~\cite{DBLP:conf/isca/HongK09} describe an 
analytical model that estimates the execution time of GPU kernels. 
They use this model
to identify the performance bottlenecks (via static analysis and emulation) 
and optimize the performance of a given {\cuda} kernel.
Baghsorkhi et al~\cite{Baghsorkhi:2010:APM} present a dynamically
adaptive approach for modeling performance of GPU programs 
which relies on dynamic instrumentation of the program
at the cost of adding runtime overheads. 
This model cannot statically determine the optimal program parameters 
at compile time. 
In~\cite{DBLP:conf/icpp/LimNM17}
the authors claim to have a static analysis method which does not
require running the code for determining the best configuration for
{\cuda} kernels. They analyze the assembly code generated by
the {\texttt{NVCC}} compiler and estimate the IPC 
of each instruction, but, there is no analysis of memory access pattern.
The authors assume that each {\cuda} kernel consists of a
for-loop nest and that the execution time is proportional to the input
problem size. These are, obviously, very strong assumptions, and impractical
for real world applications.
Moreover, the paper only reports on 4 test-cases 
(unlike a standard test suite such as PolyBench which includes 15 examples).

More recently, 
the author of \cite{DBLP:conf/ppopp/Volkov18,Volkov:EECS-2016-143}
has suggested a new GPU performance 
model which relies on Little's law \cite{little1961proof}, 
that is, measuring concurrency
as a product of latency and throughput.
%
The suggested model takes both warp and instruction concurrency into 
account.
This approach stands in opposition to the common view of many models, 
including those of \cite{cuda2019guide}, \cite{DBLP:conf/isca/HongK09}, \cite{DBLP:conf/ppopp/SimDKV12}, and \cite{Baghsorkhi:2010:APM},
which only consider warps as the unit of concurrency.
{
The model measures the performance in terms of 
required concurrency for the 
latency-bound cases (when occupancy is small and throughput grows with occupancy)
and throughput-bound cases
(when occupancy is large while throughput is at 
a maximum and is constant w.r.t. occupancy).
The author's analysis of models 
in 
\cite{cuda2019guide, DBLP:conf/isca/HongK09, DBLP:conf/ppopp/SimDKV12, 
	  Baghsorkhi:2010:APM}
indicates that 
the most common limitation among all of them
is the significant underestimation of occupancy
when the arithmetic intensity 
(number of arithmetic instructions per memory access)
of the code is intermediate.
}
%
Other major drawbacks include 
the low accuracy of the prediction when the code is 
memory intensive (i.e., low arithmetic intensity),
and underestimation, or overestimation, of the throughput
(in the case of \cite{Baghsorkhi:2010:APM} and \cite{DBLP:conf/micro/HuangLKL14}, respectively).
Noticeably, the author emphasizes the importance of \emph{cusp behavior};
that is, 
"hiding both arithmetic and memory latency at the same time may require more
warps than hiding either of them alone"~\cite{Volkov:EECS-2016-143}.
Cusp behavior is observed in practice,
however, it is not indicated in the previously mentioned performance prediction models.

\ifICPP
\section{Conclusions and Future Work}
\label{sec:Conclusion}
\fi


Our main objective is to provide a solution for
{\em portable performance optimization} of multithreaded programs.
In the context of {\cuda} programs, other research groups have
used techniques like
auto-tuning~\cite{grauer2012auto,Khan:2013:SAC:2400682.2400690},
dynamic instrumentation~\cite{kistler2003continuous}, or
both~\cite{song2015automated}.
Auto-tuning techniques have achieved great results
in projects such as 
ATLAS~\cite{DBLP:conf/ppsc/WhaleyD99},
FFTW~\cite{DBLP:conf/icassp/FrigoJ98}, and
SPIRAL~\cite{DBLP:journals/ijhpca/PuschelMSXJPVJ04}
in which part of the optimization process 
occurs \textit{off-line} and then it is applied and refined \textit{on-line}.
In contrast, we propose 
to build a completely off-line tool, namely, the rational program.
%

For this purpose, we present an end-to-end novel technique for compile-time
estimation of program parameters for optimizing a 
chosen performance metric, 
targeting manycore accelerators, particularly, {\cuda} kernels.
Our method can operate without precisely knowing
the numerical values 
of any of the data or program parameters 
when the optimization code is being constructed.
This naturally leads to a case discussion depending on the values of
those parameters, which is precisely what can be accomplished by our
notion of a rational program.
\ifPATENT
	Our technique takes a holistic approach with respect 
	to hardware architectures, low level performance
	metrics, and program and data parameters. This is the first
	ever work to employ the principle that high-level performance
	metrics are piece-wise rational functions (PRFs) of
	lower-level metrics, which are themselves PRFs of program,
	hardware, and data parameters.
\fi
	Moreover, our idea is verifiably extensible to any architectural additions,
	owing to the principle that low level performance metrics
	are piece-wise rational functions of data parameters, 
	hardware parameters, and program parameters.
	Our approach relies on the instrumentation of the program\footnote{We observe 
        that instrumentation (either via profiling or based on emulation)
        is unavoidable unless the multithreaded kernel to be optimized
        has a very specific structure, say it simply
        consists of a static for-loop nest. 
        This observation is an immediate consequence of the fact
        that, for Turing machines, the halting problem is undecidable.}, 
    and collection 
	of certain runtime values for few selected small input data sizes,
	to approximate the runtime
	performance characteristics of the program at larger input
	data sizes. Program instrumentations,
        in conjunction with source code analysis and theoretical 
        modeling, are central to avoid costly dynamic analysis.
       Consequently, 
       compilers will be able to tune applications for the target hardware 
       at compile-time, rather than having to offload that responsibility 
       to the end users.

{
As it is observed from Tables~\ref{tab:sanitycheck_nvprof} and \ref{tab:proofofconcept_nvprof},
our current implementation suffers from inaccuracy for certain examples.
We attribute this problem to usage of the MWP-CWP 
of \cite{DBLP:conf/isca/HongK09, DBLP:conf/ppopp/SimDKV12}
as our performance prediction model.
In future work, we expect that by switching 
to the model of \cite{DBLP:conf/ppopp/Volkov18,Volkov:EECS-2016-143}, 
we will be able to improve the accuracy of the rational program estimation.
}

\ifPATENT
\input{patent}
\fi

\subsection*{Acknowledgements}
The authors would like to thank IBM Canada Ltd (CAS project 880) and
NSERC of Canada (CRD grant CRDPJ500717-16).

%
%

\bibliographystyle{ACM-Reference-Format}
\bibliography{CRD,multi+arbitrary-precision,BPAS,patents,gpu_performance_models}


\begin{thebibliography}{40}


\ifx \showCODEN    \undefined \def \showCODEN     #1{\unskip}     \fi
\ifx \showDOI      \undefined \def \showDOI       #1{#1}\fi
\ifx \showISBNx    \undefined \def \showISBNx     #1{\unskip}     \fi
\ifx \showISBNxiii \undefined \def \showISBNxiii  #1{\unskip}     \fi
\ifx \showISSN     \undefined \def \showISSN      #1{\unskip}     \fi
\ifx \showLCCN     \undefined \def \showLCCN      #1{\unskip}     \fi
\ifx \shownote     \undefined \def \shownote      #1{#1}          \fi
\ifx \showarticletitle \undefined \def \showarticletitle #1{#1}   \fi
\ifx \showURL      \undefined \def \showURL       {\relax}        \fi
\providecommand\bibfield[2]{#2}
\providecommand\bibinfo[2]{#2}
\providecommand\natexlab[1]{#1}
\providecommand\showeprint[2][]{arXiv:#2}

\bibitem[\protect\citeauthoryear{??}{cud}{2018a}]%
        {cuda2015}
 \bibinfo{year}{2018}\natexlab{a}.
\newblock \bibinfo{title}{{CUDA} Runtime {API}: v10.0}.
\newblock \bibinfo{howpublished}{{NVIDIA Corporation}}.
\newblock
\newblock
\shownote{\url{http://docs.nvidia.com/cuda/pdf/CUDA_Runtime_API.pdf}.}


\bibitem[\protect\citeauthoryear{??}{cud}{2018b}]%
        {cudatoolkit2018}
 \bibinfo{year}{2018}\natexlab{b}.
\newblock \bibinfo{title}{{CUDA} Toolkit Documentation, version 10.0}.
\newblock \bibinfo{howpublished}{{NVIDIA Corporation}}.
\newblock
\newblock
\shownote{\url{https://docs.nvidia.com/cuda/}.}


\bibitem[\protect\citeauthoryear{Aho, Sethi, and Ullman}{Aho
  et~al\mbox{.}}{1986}]%
        {Aho:1986:CPT:6448}
\bibfield{author}{\bibinfo{person}{Alfred~V. Aho}, \bibinfo{person}{Ravi
  Sethi}, {and} \bibinfo{person}{Jeffrey~D. Ullman}.}
  \bibinfo{year}{1986}\natexlab{}.
\newblock \bibinfo{booktitle}{\emph{Compilers: Principles, Techniques, and
  Tools}}.
\newblock \bibinfo{publisher}{Addison-Wesley Longman Publishing Co., Inc.},
  \bibinfo{address}{Boston, MA, USA}.
\newblock
\showISBNx{0-201-10088-6}


\bibitem[\protect\citeauthoryear{Anderson, Bai, Bischof, Blackford, Demmel,
  Dongarra, Du~Croz, Greenbaum, Hammarling, McKenney, and Sorensen}{Anderson
  et~al\mbox{.}}{1999}]%
        {userguide:lapack}
\bibfield{author}{\bibinfo{person}{E. Anderson}, \bibinfo{person}{Z. Bai},
  \bibinfo{person}{C. Bischof}, \bibinfo{person}{S. Blackford},
  \bibinfo{person}{J. Demmel}, \bibinfo{person}{J. Dongarra},
  \bibinfo{person}{J. Du~Croz}, \bibinfo{person}{A. Greenbaum},
  \bibinfo{person}{S. Hammarling}, \bibinfo{person}{A. McKenney}, {and}
  \bibinfo{person}{D. Sorensen}.} \bibinfo{year}{1999}\natexlab{}.
\newblock \bibinfo{booktitle}{\emph{{LAPACK} Users' Guide}
  (\bibinfo{edition}{third} ed.)}.
\newblock \bibinfo{publisher}{Society for Industrial and Applied Mathematics},
  \bibinfo{address}{Philadelphia, PA}.
\newblock
\showISBNx{0-89871-447-8 (paperback)}


\bibitem[\protect\citeauthoryear{Asadi, Brandt, Chen, Covanov, Mansouri,
  Mohajerani, Moir, {Moreno Maza}, Xie, and Xie}{Asadi et~al\mbox{.}}{2018}]%
        {bpasweb}
\bibfield{author}{\bibinfo{person}{Mohammadali Asadi},
  \bibinfo{person}{Alexander Brandt}, \bibinfo{person}{Changbo Chen},
  \bibinfo{person}{Svyatoslav Covanov}, \bibinfo{person}{Farnam Mansouri},
  \bibinfo{person}{Davood Mohajerani}, \bibinfo{person}{Robert Moir},
  \bibinfo{person}{Marc {Moreno Maza}}, \bibinfo{person}{Ning Xie}, {and}
  \bibinfo{person}{Yuzhen Xie}.} \bibinfo{year}{2018}\natexlab{}.
\newblock \bibinfo{title}{{B}asic {P}olynomial {A}lgebra {S}ubprograms
  ({BPAS})}.
\newblock
\newblock
\newblock
\shownote{\url{http://www.bpaslib.org}.}


\bibitem[\protect\citeauthoryear{Baghsorkhi, Delahaye, Patel, Gropp, and
  Hwu}{Baghsorkhi et~al\mbox{.}}{2010}]%
        {Baghsorkhi:2010:APM}
\bibfield{author}{\bibinfo{person}{Sara~S. Baghsorkhi},
  \bibinfo{person}{Matthieu Delahaye}, \bibinfo{person}{Sanjay~J. Patel},
  \bibinfo{person}{William~D. Gropp}, {and} \bibinfo{person}{Wen-mei~W. Hwu}.}
  \bibinfo{year}{2010}\natexlab{}.
\newblock \showarticletitle{An Adaptive Performance Modeling Tool for GPU
  Architectures}. In \bibinfo{booktitle}{\emph{Proceedings of the 15th ACM
  SIGPLAN Symposium on Principles and Practice of Parallel Programming}}
  \emph{(\bibinfo{series}{PPoPP '10})}. \bibinfo{publisher}{ACM},
  \bibinfo{pages}{105--114}.
\newblock
\showISBNx{978-1-60558-877-3}


\bibitem[\protect\citeauthoryear{Beckermann}{Beckermann}{2000}]%
        {beckermann2000condition}
\bibfield{author}{\bibinfo{person}{Bernhard Beckermann}.}
  \bibinfo{year}{2000}\natexlab{}.
\newblock \showarticletitle{The condition number of real Vandermonde, Krylov
  and positive definite Hankel matrices}.
\newblock \bibinfo{journal}{\emph{Numer. Math.}} \bibinfo{volume}{85},
  \bibinfo{number}{4} (\bibinfo{year}{2000}), \bibinfo{pages}{553--577}.
\newblock


\bibitem[\protect\citeauthoryear{Brandt}{Brandt}{2018}]%
        {brandt2018high}
\bibfield{author}{\bibinfo{person}{Alexander Brandt}.}
  \bibinfo{year}{2018}\natexlab{}.
\newblock \emph{\bibinfo{title}{High Performance Sparse Multivariate
  Polynomials: Fundamental Data Structures and Algorithms}}.
\newblock \bibinfo{thesistype}{Master's\ thesis}. \bibinfo{school}{University
  of Western Ontario}, \bibinfo{address}{London, ON, Canada}.
\newblock


\bibitem[\protect\citeauthoryear{Chung and Yao}{Chung and Yao}{1977}]%
        {chung1977lattices}
\bibfield{author}{\bibinfo{person}{Kwok~Chiu Chung} {and}
  \bibinfo{person}{Te~Hai Yao}.} \bibinfo{year}{1977}\natexlab{}.
\newblock \showarticletitle{On lattices admitting unique Lagrange
  interpolations}.
\newblock \bibinfo{journal}{\emph{SIAM J. Numer. Anal.}} \bibinfo{volume}{14},
  \bibinfo{number}{4} (\bibinfo{year}{1977}), \bibinfo{pages}{735--743}.
\newblock


\bibitem[\protect\citeauthoryear{Corless and Fillion}{Corless and
  Fillion}{2013}]%
        {corless2013graduate}
\bibfield{author}{\bibinfo{person}{Robert~M Corless} {and}
  \bibinfo{person}{Nicolas Fillion}.} \bibinfo{year}{2013}\natexlab{}.
\newblock \bibinfo{booktitle}{\emph{A graduate introduction to numerical
  methods}}.
\newblock \bibinfo{publisher}{Springer}.
\newblock


\bibitem[\protect\citeauthoryear{Corporation}{Corporation}{2019a}]%
        {cuda2019guide}
\bibfield{author}{\bibinfo{person}{NVIDIA Corporation}.}
  \bibinfo{year}{2019}\natexlab{a}.
\newblock \bibinfo{title}{{CUDA C Programming Guide, v10.1.105}}.
\newblock
\newblock
\newblock
\shownote{\url{{https://docs.nvidia.com/cuda/cuda-c-programming-guide/index.html}}.}


\bibitem[\protect\citeauthoryear{Corporation}{Corporation}{2019b}]%
        {nvprofguide}
\bibfield{author}{\bibinfo{person}{NVIDIA Corporation}.}
  \bibinfo{year}{2019}\natexlab{b}.
\newblock \bibinfo{title}{{Profiler User's Guide, v10.1.105}}.
\newblock
\newblock
\newblock
\shownote{\url{https://docs.nvidia.com/cuda/profiler-users-guide/index.html}.}


\bibitem[\protect\citeauthoryear{Diamos, Kerr, Yalamanchili, and Clark}{Diamos
  et~al\mbox{.}}{2010}]%
        {DBLP:conf/IEEEpact/DiamosKYC10}
\bibfield{author}{\bibinfo{person}{Gregory~Frederick Diamos},
  \bibinfo{person}{Andrew Kerr}, \bibinfo{person}{Sudhakar Yalamanchili}, {and}
  \bibinfo{person}{Nathan Clark}.} \bibinfo{year}{2010}\natexlab{}.
\newblock \showarticletitle{Ocelot: a dynamic optimization framework for
  bulk-synchronous applications in heterogeneous systems}. In
  \bibinfo{booktitle}{\emph{19th International Conference on Parallel
  Architecture and Compilation Techniques, {PACT} 2010, Vienna, Austria,
  September 11-15, 2010}}, \bibfield{editor}{\bibinfo{person}{Valentina
  Salapura}, \bibinfo{person}{Michael Gschwind}, {and} \bibinfo{person}{Jens
  Knoop}} (Eds.). \bibinfo{publisher}{{ACM}}, \bibinfo{pages}{353--364}.
\newblock


\bibitem[\protect\citeauthoryear{Eisenbud}{Eisenbud}{2013}]%
        {eisenbud2013commutative}
\bibfield{author}{\bibinfo{person}{David Eisenbud}.}
  \bibinfo{year}{2013}\natexlab{}.
\newblock \bibinfo{booktitle}{\emph{Commutative Algebra: with a view toward
  algebraic geometry}}. Vol.~\bibinfo{volume}{150}.
\newblock \bibinfo{publisher}{Springer Science \& Business Media}.
\newblock


\bibitem[\protect\citeauthoryear{Frigo and Johnson}{Frigo and Johnson}{1998}]%
        {DBLP:conf/icassp/FrigoJ98}
\bibfield{author}{\bibinfo{person}{Matteo Frigo} {and}
  \bibinfo{person}{Steven~G. Johnson}.} \bibinfo{year}{1998}\natexlab{}.
\newblock \showarticletitle{{FFTW:} an adaptive software architecture for the
  {FFT}}. In \bibinfo{booktitle}{\emph{Proceedings of the 1998 {IEEE}
  International Conference on Acoustics, Speech and Signal Processing, {ICASSP}
  '98, Seattle, Washington, USA, May 12-15, 1998}}.
  \bibinfo{publisher}{{IEEE}}, \bibinfo{pages}{1381--1384}.
\newblock
\showISBNx{0-7803-4428-6}


\bibitem[\protect\citeauthoryear{Gibbons}{Gibbons}{1989}]%
        {Gibbons:1989:MPP:72935.72953}
\bibfield{author}{\bibinfo{person}{P.~B. Gibbons}.}
  \bibinfo{year}{1989}\natexlab{}.
\newblock \showarticletitle{A more practical {PRAM} model}. In
  \bibinfo{booktitle}{\emph{Proceedings of the ACM Symposium on Parallel
  Algorithms and Architectures}}. \bibinfo{publisher}{ACM},
  \bibinfo{pages}{158--168}.
\newblock


\bibitem[\protect\citeauthoryear{Grauer-Gray and Pouchet}{Grauer-Gray and
  Pouchet}{2012}]%
        {polybenchGPUweb}
\bibfield{author}{\bibinfo{person}{Scott Grauer-Gray} {and}
  \bibinfo{person}{Louis-Noel Pouchet}.} \bibinfo{year}{2012}\natexlab{}.
\newblock \bibinfo{title}{Implementation of Polybench codes GPU processing}.
\newblock
\newblock
\newblock
\shownote{\url{http://web.cse.ohio-state.edu/~pouchet.2/software/polybench/GPU/index.html}.}


\bibitem[\protect\citeauthoryear{Grauer-Gray, Xu, Searles, Ayalasomayajula, and
  Cavazos}{Grauer-Gray et~al\mbox{.}}{2012}]%
        {grauer2012auto}
\bibfield{author}{\bibinfo{person}{Scott Grauer-Gray}, \bibinfo{person}{Lifan
  Xu}, \bibinfo{person}{Robert Searles}, \bibinfo{person}{Sudhee
  Ayalasomayajula}, {and} \bibinfo{person}{John Cavazos}.}
  \bibinfo{year}{2012}\natexlab{}.
\newblock \showarticletitle{Auto-tuning a high-level language targeted to {GPU}
  codes}. In \bibinfo{booktitle}{\emph{Innovative Parallel Computing (InPar),
  2012}}. IEEE, \bibinfo{pages}{1--10}.
\newblock


\bibitem[\protect\citeauthoryear{Haque, {Moreno Maza}, and Xie}{Haque
  et~al\mbox{.}}{2015}]%
        {DBLP:conf/parco/HaqueMX15}
\bibfield{author}{\bibinfo{person}{Sardar~Anisul Haque}, \bibinfo{person}{Marc
  {Moreno Maza}}, {and} \bibinfo{person}{Ning Xie}.}
  \bibinfo{year}{2015}\natexlab{}.
\newblock \showarticletitle{A Many-Core Machine Model for Designing Algorithms
  with Minimum Parallelism Overheads}. In \bibinfo{booktitle}{\emph{Parallel
  Computing: On the Road to Exascale, Proceedings of the International
  Conference on Parallel Computing, ParCo 2015}}
  \emph{(\bibinfo{series}{Advances in Parallel Computing})},
  Vol.~\bibinfo{volume}{27}. \bibinfo{publisher}{{IOS} Press},
  \bibinfo{pages}{35--44}.
\newblock


\bibitem[\protect\citeauthoryear{Hong}{Hong}{1990}]%
        {hong90}
\bibfield{author}{\bibinfo{person}{H. Hong}.} \bibinfo{year}{1990}\natexlab{}.
\newblock \showarticletitle{An improvement of the projection operator in
  cylindrical algebraic decomposition}. In \bibinfo{booktitle}{\emph{ISSAC
  '90}}. \bibinfo{publisher}{ACM}, \bibinfo{pages}{261--264}.
\newblock


\bibitem[\protect\citeauthoryear{Hong and Kim}{Hong and Kim}{2009}]%
        {DBLP:conf/isca/HongK09}
\bibfield{author}{\bibinfo{person}{Sunpyo Hong} {and} \bibinfo{person}{Hyesoon
  Kim}.} \bibinfo{year}{2009}\natexlab{}.
\newblock \showarticletitle{An analytical model for a {GPU} architecture with
  memory-level and thread-level parallelism awareness}. In
  \bibinfo{booktitle}{\emph{36th International Symposium on Computer
  Architecture {(ISCA} 2009)}}. \bibinfo{pages}{152--163}.
\newblock


\bibitem[\protect\citeauthoryear{Huang, Lee, Kim, and Lee}{Huang
  et~al\mbox{.}}{2014}]%
        {DBLP:conf/micro/HuangLKL14}
\bibfield{author}{\bibinfo{person}{Jen{-}Cheng Huang},
  \bibinfo{person}{Joo~Hwan Lee}, \bibinfo{person}{Hyesoon Kim}, {and}
  \bibinfo{person}{Hsien{-}Hsin~S. Lee}.} \bibinfo{year}{2014}\natexlab{}.
\newblock \showarticletitle{GPUMech: {GPU} Performance Modeling Technique Based
  on Interval Analysis}. In \bibinfo{booktitle}{\emph{47th Annual {IEEE/ACM}
  International Symposium on Microarchitecture, {MICRO} 2014, Cambridge, United
  Kingdom, December 13-17, 2014}}. \bibinfo{publisher}{{IEEE} Computer
  Society}, \bibinfo{pages}{268--279}.
\newblock
\showISBNx{978-1-4799-6998-2}


\bibitem[\protect\citeauthoryear{Khan, Basu, Rudy, Hall, Chen, and Chame}{Khan
  et~al\mbox{.}}{2013}]%
        {Khan:2013:SAC:2400682.2400690}
\bibfield{author}{\bibinfo{person}{Malik Khan}, \bibinfo{person}{Protonu Basu},
  \bibinfo{person}{Gabe Rudy}, \bibinfo{person}{Mary Hall},
  \bibinfo{person}{Chun Chen}, {and} \bibinfo{person}{Jacqueline Chame}.}
  \bibinfo{year}{2013}\natexlab{}.
\newblock \showarticletitle{A Script-based Autotuning Compiler System to
  Generate High-performance {CUDA} Code}.
\newblock \bibinfo{journal}{\emph{ACM Trans. Archit. Code Optim.}}
  \bibinfo{volume}{9}, \bibinfo{number}{4}, Article \bibinfo{articleno}{31}
  (\bibinfo{date}{Jan.} \bibinfo{year}{2013}), \bibinfo{numpages}{25}~pages.
\newblock
\showISSN{1544-3566}


\bibitem[\protect\citeauthoryear{Kistler and Franz}{Kistler and Franz}{2003}]%
        {kistler2003continuous}
\bibfield{author}{\bibinfo{person}{Thomas Kistler} {and}
  \bibinfo{person}{Michael Franz}.} \bibinfo{year}{2003}\natexlab{}.
\newblock \showarticletitle{Continuous program optimization: A case study}.
\newblock \bibinfo{journal}{\emph{ACM Transactions on Programming Languages and
  Systems (TOPLAS)}} \bibinfo{volume}{25}, \bibinfo{number}{4}
  (\bibinfo{year}{2003}), \bibinfo{pages}{500--548}.
\newblock


\bibitem[\protect\citeauthoryear{Lim, Norris, and Malony}{Lim
  et~al\mbox{.}}{2017}]%
        {DBLP:conf/icpp/LimNM17}
\bibfield{author}{\bibinfo{person}{Robert~V. Lim}, \bibinfo{person}{Boyana
  Norris}, {and} \bibinfo{person}{Allen~D. Malony}.}
  \bibinfo{year}{2017}\natexlab{}.
\newblock \showarticletitle{Autotuning {GPU} Kernels via Static and Predictive
  Analysis}. In \bibinfo{booktitle}{\emph{46th International Conference on
  Parallel Processing, {ICPP} 2017, Bristol, United Kingdom, August 14-17,
  2017}}. \bibinfo{pages}{523--532}.
\newblock


\bibitem[\protect\citeauthoryear{Little}{Little}{1961}]%
        {little1961proof}
\bibfield{author}{\bibinfo{person}{John~DC Little}.}
  \bibinfo{year}{1961}\natexlab{}.
\newblock \showarticletitle{A proof for the queuing formula: L= $\lambda$ W}.
\newblock \bibinfo{journal}{\emph{Operations research}} \bibinfo{volume}{9},
  \bibinfo{number}{3} (\bibinfo{year}{1961}), \bibinfo{pages}{383--387}.
\newblock


\bibitem[\protect\citeauthoryear{Ma, Agrawal, and Chamberlain}{Ma
  et~al\mbox{.}}{2014}]%
        {ma2014memory}
\bibfield{author}{\bibinfo{person}{L. Ma}, \bibinfo{person}{K. Agrawal}, {and}
  \bibinfo{person}{R.~D. Chamberlain}.} \bibinfo{year}{2014}\natexlab{}.
\newblock \showarticletitle{A memory access model for highly-threaded many-core
  architectures}.
\newblock \bibinfo{journal}{\emph{Future Generation Computer Systems}}
  \bibinfo{volume}{30} (\bibinfo{year}{2014}), \bibinfo{pages}{202--215}.
\newblock


\bibitem[\protect\citeauthoryear{Mei and Chu}{Mei and Chu}{2017}]%
        {DBLP:journals/tpds/MeiC17}
\bibfield{author}{\bibinfo{person}{Xinxin Mei} {and} \bibinfo{person}{Xiaowen
  Chu}.} \bibinfo{year}{2017}\natexlab{}.
\newblock \showarticletitle{Dissecting {GPU} Memory Hierarchy Through
  Microbenchmarking}.
\newblock \bibinfo{journal}{\emph{{IEEE} Trans. Parallel Distrib. Syst.}}
  \bibinfo{volume}{28}, \bibinfo{number}{1} (\bibinfo{year}{2017}),
  \bibinfo{pages}{72--86}.
\newblock


\bibitem[\protect\citeauthoryear{Nickolls, Buck, Garland, and Skadron}{Nickolls
  et~al\mbox{.}}{2008}]%
        {Nickolls:2008:SPP:1365490.1365500}
\bibfield{author}{\bibinfo{person}{J. Nickolls}, \bibinfo{person}{I. Buck},
  \bibinfo{person}{M. Garland}, {and} \bibinfo{person}{K. Skadron}.}
  \bibinfo{year}{2008}\natexlab{}.
\newblock \showarticletitle{Scalable Parallel Programming with {CUDA}}.
\newblock \bibinfo{journal}{\emph{Queue}} \bibinfo{volume}{6},
  \bibinfo{number}{2} (\bibinfo{year}{2008}), \bibinfo{pages}{40--53}.
\newblock
\showISSN{1542-7730}


\bibitem[\protect\citeauthoryear{Olver}{Olver}{2006}]%
        {olver2006multivariate}
\bibfield{author}{\bibinfo{person}{Peter~J Olver}.}
  \bibinfo{year}{2006}\natexlab{}.
\newblock \showarticletitle{On multivariate interpolation}.
\newblock \bibinfo{journal}{\emph{Studies in Applied Mathematics}}
  \bibinfo{volume}{116}, \bibinfo{number}{2} (\bibinfo{year}{2006}),
  \bibinfo{pages}{201--240}.
\newblock


\bibitem[\protect\citeauthoryear{P{\"{u}}schel, Moura, Singer, Xiong, Johnson,
  Padua, Veloso, and Johnson}{P{\"{u}}schel et~al\mbox{.}}{2004}]%
        {DBLP:journals/ijhpca/PuschelMSXJPVJ04}
\bibfield{author}{\bibinfo{person}{Markus P{\"{u}}schel},
  \bibinfo{person}{Jos{\'{e}} M.~F. Moura}, \bibinfo{person}{Bryan Singer},
  \bibinfo{person}{Jianxin Xiong}, \bibinfo{person}{Jeremy~R. Johnson},
  \bibinfo{person}{David~A. Padua}, \bibinfo{person}{Manuela~M. Veloso}, {and}
  \bibinfo{person}{Robert~W. Johnson}.} \bibinfo{year}{2004}\natexlab{}.
\newblock \showarticletitle{Spiral: {A} Generator for Platform-Adapted
  Libraries of Signal Processing Alogorithms}.
\newblock \bibinfo{journal}{\emph{{IJHPCA}}} \bibinfo{volume}{18},
  \bibinfo{number}{1} (\bibinfo{year}{2004}), \bibinfo{pages}{21--45}.
\newblock


\bibitem[\protect\citeauthoryear{Ryoo, Rodrigues, Baghsorkhi, Stone, Kirk, and
  Hwu}{Ryoo et~al\mbox{.}}{2008a}]%
        {ryoo2008optimization}
\bibfield{author}{\bibinfo{person}{Shane Ryoo}, \bibinfo{person}{Christopher~I
  Rodrigues}, \bibinfo{person}{Sara~S Baghsorkhi}, \bibinfo{person}{Sam~S
  Stone}, \bibinfo{person}{David~B Kirk}, {and} \bibinfo{person}{Wen-mei~W
  Hwu}.} \bibinfo{year}{2008}\natexlab{a}.
\newblock \showarticletitle{Optimization principles and application performance
  evaluation of a multithreaded GPU using CUDA}. In
  \bibinfo{booktitle}{\emph{Proceedings of the 13th ACM SIGPLAN Symposium on
  Principles and practice of parallel programming}}. ACM,
  \bibinfo{pages}{73--82}.
\newblock


\bibitem[\protect\citeauthoryear{Ryoo, Rodrigues, Stone, Baghsorkhi, Ueng,
  Stratton, and Hwu}{Ryoo et~al\mbox{.}}{2008b}]%
        {RRSBUSH2008}
\bibfield{author}{\bibinfo{person}{S. Ryoo}, \bibinfo{person}{C.~I. Rodrigues},
  \bibinfo{person}{S.~S. Stone}, \bibinfo{person}{S.~S. Baghsorkhi},
  \bibinfo{person}{S. Ueng}, \bibinfo{person}{J.~A. Stratton}, {and}
  \bibinfo{person}{W.~W. Hwu}.} \bibinfo{year}{2008}\natexlab{b}.
\newblock \showarticletitle{Program optimization space pruning for a
  multithreaded {GPU}}. In \bibinfo{booktitle}{\emph{Proc. of CGO}}.
  \bibinfo{publisher}{ACM}, \bibinfo{pages}{195--204}.
\newblock


\bibitem[\protect\citeauthoryear{Sim, Dasgupta, Kim, and Vuduc}{Sim
  et~al\mbox{.}}{2012}]%
        {DBLP:conf/ppopp/SimDKV12}
\bibfield{author}{\bibinfo{person}{Jaewoong Sim}, \bibinfo{person}{Aniruddha
  Dasgupta}, \bibinfo{person}{Hyesoon Kim}, {and} \bibinfo{person}{Richard~W.
  Vuduc}.} \bibinfo{year}{2012}\natexlab{}.
\newblock \showarticletitle{A performance analysis framework for identifying
  potential benefits in {GPGPU} applications}. In
  \bibinfo{booktitle}{\emph{Proceedings of the 17th {ACM} {SIGPLAN} Symposium
  on Principles and Practice of Parallel Programming, {PPOPP} 2012, New
  Orleans, LA, USA, February 25-29, 2012}}. \bibinfo{pages}{11--22}.
\newblock


\bibitem[\protect\citeauthoryear{Song, Wang, and Mart{\'i}nez}{Song
  et~al\mbox{.}}{2015}]%
        {song2015automated}
\bibfield{author}{\bibinfo{person}{Chenchen Song}, \bibinfo{person}{Lee-Ping
  Wang}, {and} \bibinfo{person}{Todd~J Mart{\'i}nez}.}
  \bibinfo{year}{2015}\natexlab{}.
\newblock \showarticletitle{Automated Code Engine for Graphical Processing
  Units: Application to the Effective Core Potential Integrals and Gradients}.
\newblock \bibinfo{journal}{\emph{Journal of chemical theory and computation}}
  (\bibinfo{year}{2015}).
\newblock


\bibitem[\protect\citeauthoryear{Stockmeyer and Vishkin}{Stockmeyer and
  Vishkin}{1984}]%
        {DBLP:journals/siamcomp/StockmeyerV84}
\bibfield{author}{\bibinfo{person}{L.~J. Stockmeyer} {and} \bibinfo{person}{U.
  Vishkin}.} \bibinfo{year}{1984}\natexlab{}.
\newblock \showarticletitle{Simulation of Parallel Random Access Machines by
  Circuits}.
\newblock \bibinfo{journal}{\emph{SIAM J. Comput.}} \bibinfo{volume}{13},
  \bibinfo{number}{2} (\bibinfo{year}{1984}), \bibinfo{pages}{409--422}.
\newblock


\bibitem[\protect\citeauthoryear{Volkov}{Volkov}{2016}]%
        {Volkov:EECS-2016-143}
\bibfield{author}{\bibinfo{person}{Vasily Volkov}.}
  \bibinfo{year}{2016}\natexlab{}.
\newblock \emph{\bibinfo{title}{Understanding Latency Hiding on GPUs}}.
\newblock \bibinfo{thesistype}{Ph.D. Dissertation}. \bibinfo{school}{EECS
  Department, University of California, Berkeley}.
\newblock


\bibitem[\protect\citeauthoryear{Volkov}{Volkov}{2018}]%
        {DBLP:conf/ppopp/Volkov18}
\bibfield{author}{\bibinfo{person}{Vasily Volkov}.}
  \bibinfo{year}{2018}\natexlab{}.
\newblock \showarticletitle{A microbenchmark to study {GPU} performance
  models}. In \bibinfo{booktitle}{\emph{Proceedings of the 23rd {ACM} {SIGPLAN}
  Symposium on Principles and Practice of Parallel Programming, PPoPP 2018,
  Vienna, Austria, February 24-28, 2018}},
  \bibfield{editor}{\bibinfo{person}{Andreas Krall} {and}
  \bibinfo{person}{Thomas~R. Gross}} (Eds.). \bibinfo{publisher}{{ACM}},
  \bibinfo{pages}{421--422}.
\newblock


\bibitem[\protect\citeauthoryear{Whaley and Dongarra}{Whaley and
  Dongarra}{1998}]%
        {DBLP:conf/ppsc/WhaleyD99}
\bibfield{author}{\bibinfo{person}{R.~Clinton Whaley} {and}
  \bibinfo{person}{Jack Dongarra}.} \bibinfo{year}{1998}\natexlab{}.
\newblock \showarticletitle{Automatically Tuned Linear Algebra Software}. In
  \bibinfo{booktitle}{\emph{Proceedings of the 1998 IEEE/ACM Conference on
  Supercomputing}}.
\newblock


\bibitem[\protect\citeauthoryear{Wong, Papadopoulou, Sadooghi{-}Alvandi, and
  Moshovos}{Wong et~al\mbox{.}}{2010}]%
        {DBLP:conf/ispass/WongPSM10}
\bibfield{author}{\bibinfo{person}{Henry Wong}, \bibinfo{person}{Misel{-}Myrto
  Papadopoulou}, \bibinfo{person}{Maryam Sadooghi{-}Alvandi}, {and}
  \bibinfo{person}{Andreas Moshovos}.} \bibinfo{year}{2010}\natexlab{}.
\newblock \showarticletitle{Demystifying {GPU} microarchitecture through
  microbenchmarking}. In \bibinfo{booktitle}{\emph{{IEEE} International
  Symposium on Performance Analysis of Systems and Software, {ISPASS} 2010,
  28-30 March 2010, White Plains, NY, {USA}}}. \bibinfo{publisher}{{IEEE}
  Computer Society}, \bibinfo{pages}{235--246}.
\newblock
\showISBNx{978-1-4244-6022-9}


\end{thebibliography}
\end{document}